\documentclass[a4paper,12pt]{article}
\ifx\pdfoutput\undefined
\usepackage[dvips,bookmarks=false]{hyperref}	% This is for arXiv.org
\else
\usepackage{hyperref}	% This is for pdftex
\fi
\hypersetup{colorlinks,bookmarksopen,bookmarksnumbered,citecolor=blue,
linkcolor=black,pdfstartview=FitH,urlcolor=blue}

\usepackage{amsmath}
\usepackage{amssymb}
\usepackage{cite}
\usepackage{braket}
\usepackage{bm}
\usepackage{color}
\usepackage{hhline}
\usepackage{multirow}
\usepackage{indentfirst}
\usepackage{graphicx}

\allowdisplaybreaks

\newcommand{\Slash}[1]{{\ooalign{\hfil/\hfil\crcr$#1$}}}

\renewcommand{\bar}{\overline}

\if0
\usepackage[dvipdfmx]{hyperref}
\usepackage{pxjahyper} %%after hyperref
\hypersetup{colorlinks,bookmarksopen,bookmarksnumbered,citecolor=blue,
linkcolor=black,pdfstartview=FitH,urlcolor=blue}
\fi

\usepackage{ulem}
\begin{document}

%-==========-
% Title page
%-==========-
\begin{titlepage}

\begin{flushright}
KUNS-2811
\end{flushright}

\begin{center}

\vspace{1cm}
{\large\textbf{
TeV-scale Majorogenesis
}
 }
\vspace{1cm}

\renewcommand{\thefootnote}{\fnsymbol{footnote}}
Yoshihiko Abe\footnote[1]{y.abe@gauge.scphys.kyoto-u.ac.jp}
,
Yu Hamada\footnote[2]{yu.hamada@gauge.scphys.kyoto-u.ac.jp}
,
Takahiro Ohata\footnote[3]{tk.ohata@gauge.scphys.kyoto-u.ac.jp}
,
\\
Kenta Suzuki\footnote[4]{S.Kenta@gauge.scphys.kyoto-u.ac.jp}
,
Koichi Yoshioka\footnote[5]{yoshioka@gauge.scphys.kyoto-u.ac.jp}
\vspace{5mm}

\textit{%
{Department of Physics, Kyoto University, Kyoto 606-8502, Japan}\\
}

\vspace{8mm}

\abstract{%-----
The Majoron, the Nambu-Goldstone boson of lepton number symmetry, is an interesting candidate for dark matter as it deeply connects the dark matter and neutrino physics.
In this paper, 
we consider the Majoron dark matter as pseudo Nambu-Goldstone boson with TeV-scale mass.
The heavy Majoron generally has the large decay constant and tiny Yukawa couplings to light right-handed neutrinos which are required by cosmological and astrophysical observations.
That makes it difficult to realize the desired amount of the relic abundance of Majoron dark matter.
We consider three improved scenarios for the generation of Majoron, dubbed as Majorogenesis, in the early universe and find in all cases the parameter space compatible with the relic abundance and cosmic-ray constraints.
}%-----

\end{center}
\end{titlepage}

\renewcommand{\thefootnote}{\arabic{footnote}}
\newcommand{\bhline}[1]{\noalign{\hrule height #1}}
\newcommand{\bvline}[1]{\vrule width #1}

\setcounter{footnote}{0}

\setcounter{page}{1}
%%%%%%%%%%%%%%%%%%%%%%%%%%%%%%%%%%%%%%

\tableofcontents

%%%%%%%%%%%%%%%
\section{Introduction}
\label{sec:1}

The existence of dark matter(DM) is clear from various observations over the past decades,
such as galaxy rotation curves\cite{Corbelli:1999af,Sofue:2000jx}, gravitational lensing\cite{Massey:2010hh}, cosmic microwave background\cite{Aghanim:2018eyx} and collision of Bullet Cluster\cite{Randall:2007ph}.
While there are various constraints on the DM mass and the scattering cross section
from astrophysical observations and direct detection experiments\cite{Ackermann:2013yva,Sevilla:2005wx,Akerib:2017kat,Cui:2017nnn,Aprile:2018dbl},
the nature of DM is still unknown.
The identification of DM is important not only for cosmology but also for particle physics, 
because there are no particle contents playing the role of DM in the Standard Model(SM).
DM would be the key to investigate new physics beyond the SM.
Another important issue that is unanswered by the SM is the neutrino masses\cite{Tanabashi:2018oca},
which is implied by the observations of neutrino oscillation.
One way to realize the tiny neutrino mass scale is the see-saw mechanism
\cite{Minkowski:1977sc,Yanagida:1980xy,GellMann:1980vs}.
In the type-I see-saw, right-handed(RH) neutrinos and their Majorana masses are introduced to obtain the realistic neutrino masses.
However, the origin of the Majorana masses is not explained within the model.
In the so-called Majoron model \cite{Chikashige:1980qk,Chikashige:1980ui,Gelmini:1980re},
a new SM-singlet complex scalar is introduced to explain it.
The scalar develops a vacuum expectation value(VEV) breaking a global $U(1)$ symmetry, 
which provides the Majorana masses as the SM Higgs mechanism.
Corresponding to the symmetry breaking,
there arises a pseudo scalar particle called the Majoron, 
which is the Nambu-Goldstone boson(NGB) associated with the $U(1)$ symmetry.
In Refs.~\cite{Gu:2010ys,Bazzocchi:2008fh,Berezinsky:1993fm,Lattanzi:2013uza,Gelmini:1984pe,Heeck:2019guh},
it is discussed that
the Majoron can be a DM candidate by introducing explicit breaking terms for the $U(1)$ symmetry.
In particular, the Majoron becomes a pseudo NGB(pNGB) when the soft breaking mass term is introduced as in Ref.~\cite{Gu:2010ys}.
A remarkable feature of pNGB DM is the derivative coupling with other (scalar) particles 
which enables us to avoid the constraints from the direct detection experiments
\cite{Barger:2010yn,Gross:2017dan,Azevedo:2018exj,Ishiwata:2018sdi}.
For further studies on the Majoron DM and the pNGB DM, see, e.g.,
Refs.~\cite{PalomaresRuiz:2007ry,Covi:2009xn,Matsumoto:2010hz,Queiroz:2014yna,Barducci:2016fue,Huitu:2018gbc,Cline:2019okt,Ruhdorfer:2019utl,Arina:2019tib}.
The origin of the $U(1)$ breaking term in the scalar potential is discussed in various contexts
such as the effect of quantum gravity \cite{Rothstein:1992rh},
neutrino Dirac Yukawa coupling \cite{Frigerio:2011in},
coupling with another scalar \cite{Abe:2020iph,Okada:2020zxo} and so on.
On the other hand, some cosmic-ray observations are known to suggest the existence of leptophilic TeV-scale DM\@.
That motivates us to consider a TeV-scale Majoron DM,
whose mass scale is heavier than those in previous works.
The heavy Majoron can decay to neutrinos, which requires
the SM singlet scalar VEV is around the unification scale.
It can also decay to heavy quarks such as the top quark,
and that imposes strong upper bound on the Yukawa couplings
between the Majoron and the RH neutrinos.
The Majoron interactions are too small to realize the DM relic abundance
via the thermal freeze-out mechanism \cite{Arina:2019tib}.
Hence the creation of Majoron DM (dubbed as Majorogenesis) at TeV scale should be realized in a way other than the freeze-out mechanism, such as the freeze-in production \cite{Hall:2009bx}.
In this paper, we investigate the Majorogenesis for TeV-scale Majoron.
We then consider the following three scenarios;
(A) introducing explicit Majoron masses,
(B) using the interaction with the SM Higgs doublet
(C) using the resonant production from non-thermal RH neutrinos.
All of these scenarios are found to have the parameter space compatible with
the tiny Yukawa coupling and the DM relic abundance.
This paper is organized as follows.
In Sec.~\ref{052320_27Mar20}, we discuss the Majoron model and its phenomenological constraints from heavy Majoron DM decays.
In Sec.~\ref{052339_27Mar20}, we show the difficulty of creating the heavy Majoron in the reference model,
and then consider three ways to realize the TeV-scale Majorogenesis.
In each case, we will evaluate the Majoron relic abundance and show the parameter space realizing the TeV-scale Majorogenesis.
Sec.~\ref{052401_27Mar20} is devoted to summarizing our results
and discussing future work.
%

%%%%%%%%%%%%%%%
\section{Majoron Dark Matter}
\label{052320_27Mar20}

%%%
\subsection{The model}
\label{subsec:model}

First of all, we consider the reference Majoron model for the following discussion.
We introduce a new SM-singlet complex scalar which has the Yukawa coupling to RH neutrinos.
The Lagrangian for the RH neutrinos $\nu_{Ri}$ are written as
% -----
\begin{align}
 \mathcal{L}_{N} = i \bar{\nu_{Ri}} \Slash{\partial} \nu_{Ri} -\frac{f_{ij}}{2} \Phi \bar{\nu^c_{Ri}} \nu_{Rj} - y^\nu_{\alpha i} \bar{L_\alpha} \tilde{H} \nu_{Ri} + \mathrm{h.c.},\label{155536_30Mar20}
\end{align}
% -----
where the RH neutrinos and the new scalar $\Phi$ have the lepton number $+1$ and $-2$, respectively.
The neutrino Yukawa coupling $y^{\nu}_{\alpha i}$ gives the Dirac mass $m_{D} = y^{\nu} v/\sqrt{2}$ after the electroweak symmetry breaking,
where $v$ is the electroweak VEV $v \simeq 246 ~\mathrm{GeV}$.
In addition, the new Yukawa coupling with $\Phi$ gives the Majorana mass $M_{N} = f v_{\phi} /\sqrt{2}$.
Thus, the small masses for active neutrinos are generated by the type-I seesaw mechanism as 
$(m_{\nu})_{\alpha \beta}  \approx - ( m_{D})_{\alpha i} ( M_{N}^{-1})_{ij} (m_{D}^{\mathrm{T}})_{j \beta}$.
We use Geek indices $\alpha , \beta,\ldots$ for the generation of the SM leptons and Latin indices $i, j, \ldots$ for the generation of the RH neutrinos.
The scalar potential in the model is written as
% -----
\begin{align}
 V(H, \Phi) = V_H (H)-\frac{\mu_{\Phi}^2}{2} |\Phi|^2+ \frac{\lambda_{\Phi}}{2} |\Phi|^4
 -\frac{m^2}{4} \bigl( \Phi^2 + {\Phi^*}^2 \bigr),\label{151108_26Mar20}
\end{align}
% -----
where $V_H$ is the Higgs potential in the SM
and the coupling between $\Phi$ and $H$ will be taken into account in Sec.~\ref{subsec:scalar}.
The last quadratic term proportional to $m^2$ is the soft-breaking term to generate the pNGB mass.
This term breaks the $U(1)_L$ symmetry of the scalar potential into $\mathbb{Z}_{2}$, 
which corresponds to $\Phi \to - \Phi$.\footnote{%-----
The total Lagrangian with this soft-breaking term is invariant under the $\mathbb{Z}_4$ symmetry, which is the residual discrete symmetry of the global $U(1)_L$.
}
For the potential stability, the quartic coupling satisfies $\lambda_{\Phi} >0$.
The scalar field develops a VEV $v_{\phi}$, and is parametrized as
% -----
\begin{align}
 \Phi = \frac{ v_\phi + \phi + i \chi}{\sqrt{2}}.
\end{align}
% -----
The stationary conditions are solved as $\mu_{\Phi}^2 = \lambda_{\Phi} v_{\phi}^2-m^2$, 
and the scalar masses in the $U(1)_{L}$ breaking vacuum are given by
% -----
\begin{align}
 m_{\phi}^2 = \lambda_{\Phi} v_{\phi}^2,
 \quad
 m_{\chi}^2 = m^2.
\end{align}
% -----
The CP-odd component $\chi$ is a pNGB called as the Majoron,
whose mass is given by the soft-breaking parameter $m$.
In the following parts of this paper, we will see that this Majoron can be a DM candidate.
In general, the Yukawa matrix $f_{ij}$ in Eq.~\eqref{155536_30Mar20} can be diagonalized into $f_i \delta_{ij}$ by the redefinition of the RH neutrinos and the diagonal couplings $f_{i}$ are taken to be real.
The Majorana fermion in this mass basis is denoted by $N_{i} = \nu_{Ri} + \nu^{c}_{Ri}$, 
in which we denote the redefined RH neutrino as $\nu_{Ri}$.
The Lagrangian is rewritten using these Majorana fermions as 
% -----
\begin{align}
 \mathcal{L}_{N} = & \frac{i}{2} \bar{N_i} \Slash{\partial} N_i - \frac{M_{N_i}}{2} \bar{N_i} N_i -\frac{f_{i}}{2 \sqrt{2}} \phi \bar{N_i} N_i - \frac{i f_{i}}{2 \sqrt{2}} \chi \bar{N_i} \gamma_5 N_i
 \nonumber\label{145644_26Mar20}\\
 & - Y^{\nu}_{\alpha i } \bar{L_\alpha} \tilde{H} P_R N_i + \mathrm{h.c.},
\end{align}
% -----
where $Y^{\nu}_{\alpha i}$ is the neutrino Yukawa matrix in the RH neutrino mass basis and $P_{R/L}$ is the chirality projection.
An important point is that the flavor changing off-diagonal interaction between the Majoron and the RH neutrinos
such as $\chi \bar{N_1} N_2$ disappears in the mass diagonal basis.
%

%%%
\subsection{Decaying dark matter}

In this subsection, we see features of the TeV-scale Majoron
and the phenomenological constraints as the DM candidate.
The Majoron is assumed to be lighter than the lightest RH neutrino
to prevent it from decaying into the RH neutrinos.
Otherwise, the Yukawa coupling $f$ is required to be highly suppressed and/or the VEV $v_{\phi}$ must be huge due to astrophysical constraints.
%

%
%%%
\begin{figure}[t]
\centering
\includegraphics[width=0.4\textwidth]{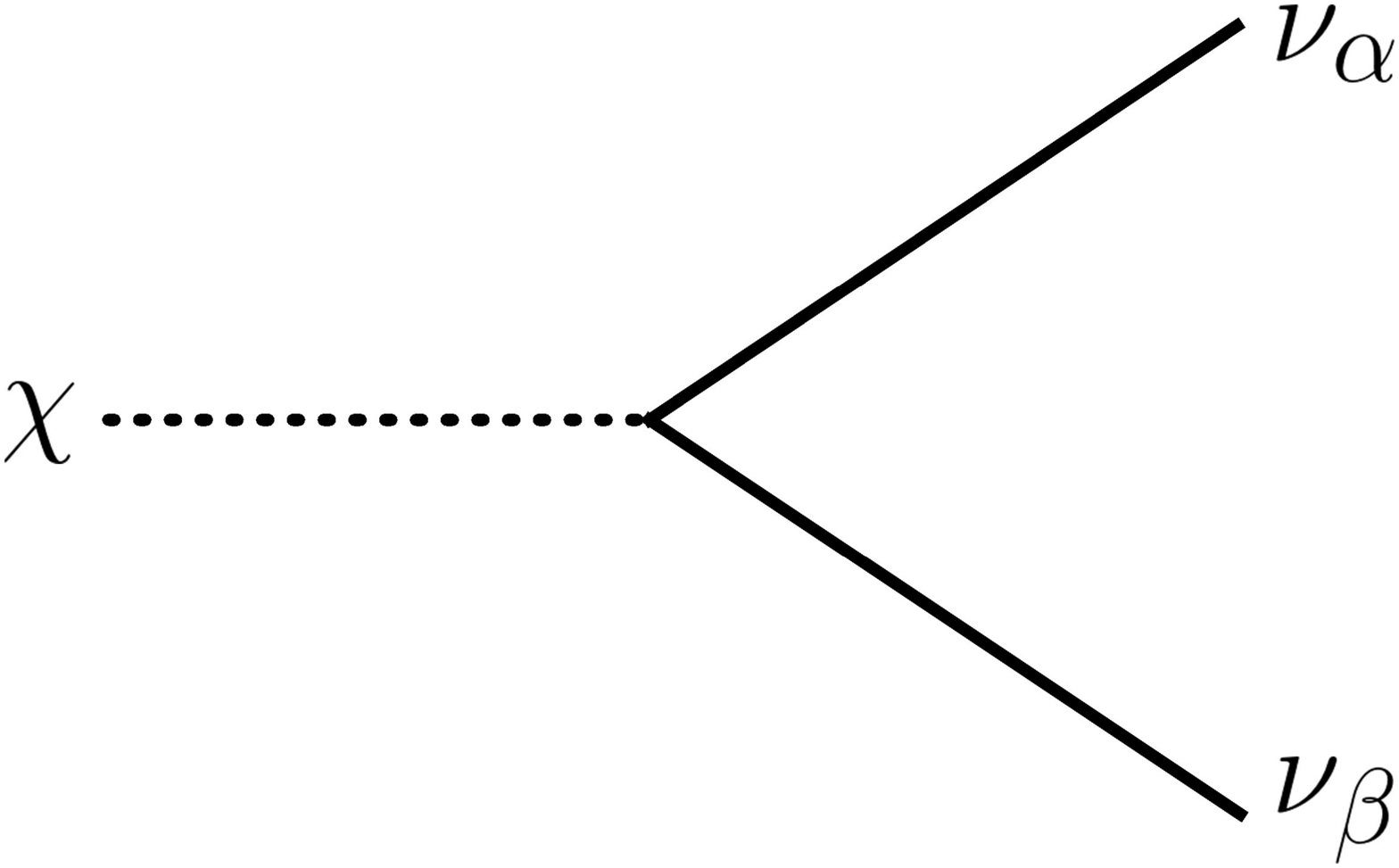}\hspace{1em}
\includegraphics[width=0.4\textwidth]{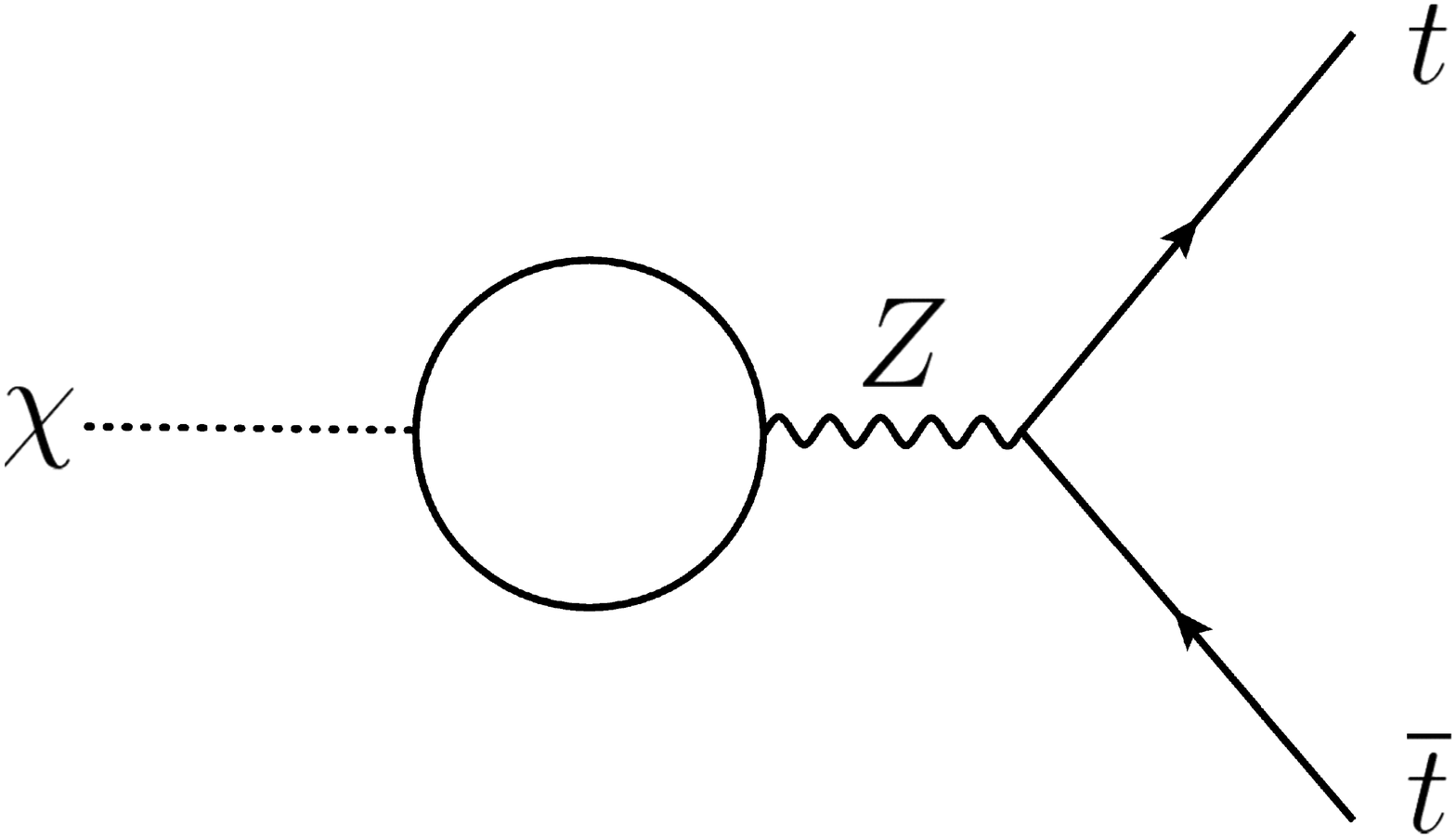}
\caption{% -----
The Feynman diagrams for the DM decay processes.
 (left): $\chi \to \nu_{\alpha} \nu_{\beta}$.
 (right): $\chi \to t \bar{t}$. The internal lines in the loop represent the active and heavy neutrinos.
}% -----
\label{fig:DMdecay}
\end{figure}
%%%
%

%
The massive Majoron is unstable due to its interaction with RH neutrinos and the neutrino Yukawa couplings.
The main decay channels are expressed by the Feynman diagrams of Fig.~\ref{fig:DMdecay}.
The decay width to the neutrinos is given by
% -----
\begin{align}
 \Gamma_{\chi \to \nu_{\alpha} \nu_{\beta}} =
 \frac{m_{\chi}}{16 \pi v_{\phi}^2} \bigl| ( m_{\nu})_{\alpha \beta} \bigr|^2,
\end{align}
% -----
where $( m_{\nu})_{\alpha \beta}$ is the neutrino mass matrix.
To realize the long-lived DM, the VEV $v_{\phi}$ has a lower bound for a fixed value of the DM mass $m_{\chi}$.
The constraints on the DM mass and lifetime for this decay mode are discussed e.g., in Refs.~\cite{PalomaresRuiz:2007ry,Covi:2009xn}.
For example, the VEV $v_{\phi}$ is found to satisfy $v_{\phi} \gtrsim 10^{15}\ \mathrm{GeV}$ for TeV-scale DM\@.
In the following parts, we assume $v_{\phi} \approx 10^{15}~\mathrm{GeV}$.
In addition, 
the Majoron is so heavy that it can decay to (the top) quark pair through the one-loop diagram shown Fig.~{\ref{fig:DMdecay}.
As the width is generally proportional to the quark mass, 
the dominant radiative decay is given by $\chi \to t \bar{t}$, if possible, 
and its width is evaluated as
% -----
\begin{align}
 \Gamma_{\chi \to t \bar{t}} = \frac{3 \alpha_{W} C_{\chi Z}^2}{8 \cos^2 \theta_{W}}
 \frac{m_{\chi} m_{t}^2}{m_{Z}^4} \sqrt{ 1 - \frac{4 m_{t}^2}{m_{\chi}^2}},
\end{align}
% -----
where $m_t$ and $m_Z$ are the masses of the top quark and the $Z$ boson, respectively, and
$\alpha_{W}$ is the fine structure constant of $SU(2)_L$ gauge coupling.
The overall factor $3$ comes from the summation of color indices of the final states.
The neutrino loop factor connecting $\chi$ and $Z$ is given by
\begin{align}
 C_{\chi Z} = \sum_{i,j}
 \frac{g \bigl| (m_D)_{ij} \bigr|^2}{ 16 \pi^2 v_\phi \cos \theta_W}
 \int_0^1 dx \int_0^{1-x} dy \int_0^{1-x-y} dz \,
 \frac{F( 2 m_\chi ^2 / M_{N_i}^2)}{F( m_\chi^2 / M_{N_i}^2)^2},
\end{align}
where $F(\omega) \equiv (x +y) + (y +z) (y+z -1) \omega $.
The main decay modes of the Majoron are these $\chi \to \nu_{\alpha} \nu_{\beta}$, $\chi \to t \bar{t}$,
and the model parameters are constrained by the cosmic-ray observations such as anti-protons and gamma-rays.
The decay widths of the Majoron to other SM particles are much smaller,
then the constraints are irrelevant.\footnote{
In the case of the Majoron being light,
see a previous work \cite{Garcia-Cely:2017oco} for the constraints from the Majoron decay.
}
In general, analyzing the constraints on the model parameters are very complicated due to many degrees of freedom and indeterminacy\cite{Iwashima:2020},
which is beyond the scope of this paper.
In this paper, we impose a conservative upper bound
on the Yukawa coupling $f_{i} \lesssim 10^{-(10\text{--}11)}$
with reference to the past analysis,
but the precise value of $f_{i}$ is irrelevant to the Majoron creation.
From these results and analysis, we find the following three statements are inseparable in the TeV-scale Majoron DM model:
%
%%%%
\begin{enumerate}
\item
Light RH neutrinos with TeV-PeV-scale masses
\item
Heavy Majoron feebly interacting with RH neutrinos
\item
Large VEV of $\Phi$ around the unification scale
\end{enumerate}
%%%
%

%
%%%%%%%%%%%%%%%
\section{Dark Matter Creation: Majorogenesis}
\label{052339_27Mar20}

In this section, we will show the difficulty to realize the DM relic abundance,
and discuss some improved scenarios for the Majorogenesis to take place.
%

%%%
\subsection{Flaw and improvements of the model}
As we have seen in the last section,
the three conditions,
1. light RH neutrinos,
2. heavy Majoron,
3. large $v_{\phi}$,
are inseparable  when we consider a TeV-scale Majoron DM.
In the model in Sec. 2, the Majoron couples to the SM particles
only through the RH neutrinos
and the coupling is too small to realize the freeze-out mechanism.
Even if we introduce the mixing coupling such as $\lambda_{H\Phi} |H|^2 |\Phi|^2$,
it is hard for pNGB DM with the large VEV to realize the relic abundance
as by the freeze-out mechanism \cite{Arina:2019tib}.
Then another option to create the Majoron is the freeze-in mechanism discussed
in Ref.~\cite{Hall:2009bx}.
The magnitude of the coupling that is necessary for the freeze-in to work is
typically $\mathcal{O}(10^{-11})$,
and thus the tiny Yukawa couplings in the model of Sec. 2,
$ f_{ij} \lesssim \mathcal{O}(10^{-(10 \text{--}11)})$,
seem useful for the Majoron creation via the freeze-in. 
However, the Yukawa interaction between the Majoron and the RH neutrinos is
flavor diagonal in the RH neutrino mass basis,
and flavor changing off-diagonal interactions such as $\chi \bar{N_{1}} N_{2}$ are
absent in the Lagrangian
(see Eq.~(\ref{145644_26Mar20})).
The other processes are too tiny to explain the relic abundance by the freeze-in mechanism.
The scattering amplitude of the annihilation $NN \to \chi\chi$ via $t$-channel is proportional to $f^2$.
In addition, the decay $N \to \chi \nu$ is highly suppressed by the neutrino mass on top of $f$.
Therefore, it is impossible to realize the DM relic abundance by the freeze-in mechanism
using the RH neutrino decay in that model.
Here let us consider the following three scenarios to avoid this flaw.
(A):
The first is to modify the universality of mass/coupling ratios for the pNGB Majoron.
A simple way for this is to introduce Majorana masses for RH neutrinos, which break the $U(1)_L$ symmetry similarly to the soft breaking term for $\Phi$.
Then flavor changing couplings of the Majoron generally appear in the RH neutrino mass basis, 
and could lead to the freeze-in production of Majoron.
(B):
The second is adding the mixing coupling between the SM Higgs $H$
and the SM singlet scalar $\Phi$ such as $\lambda_{H\Phi} |H|^2 |\Phi|^2$
to the scalar potential (\ref{151108_26Mar20}).
The Majoron can interact with the SM Higgs via this coupling on top of neutrinos,
but the typical magnitude of the interaction is also too small to realize the thermal relic
because of the nature of NGB \cite{Ruhdorfer:2019utl} as we stated above.
As an alternative option, we consider the freeze-in mechanism through this portal coupling.
(C):
The third option is using a non-thermal creation of the RH neutrinos
during the reheating after the cosmological inflation.
The scattering process mediated by the CP-even scalar particle arising from $\Phi$
is essential to explain the DM relic abundance.
In the rest part of this section, we discuss the above three scenarios (A)--(C) and
investigate the parameter space realizing the TeV-scale Majorogenesis for each case.
%

%%%
\subsection{(A): Heavy RH neutrino decay}
\label{subsec:RNdecay}

Let us consider the scenario (A), in which Majorana mass terms for the RH neutrinos are introduced:
% -----
\begin{align}
 \Delta \mathcal{L}_{\mathrm{Majorana}} = - \frac{1}{2} m_{ij} \bar{\nu^c_{Ri}} \nu_{Rj} + \mathrm{h.c.},
\end{align}
% -----
which enables the flavor changing interactions in the mass-diagonal basis.
In this subsection, we consider only two RH neutrinos ($i=1,2$),
or equivalently, we assume that one of the three is sufficiently heavy.
Hereafter, we use $g$ for the off-diagonal Yukawa interaction giving $ \chi \bar{N_{1}} N_{2}$ vertex,
which is assumed to have the constraint,
\begin{equation}
g \lesssim 10^{-10},\label{202732_26Mar20}
\end{equation}
as in the Majoron model.
%

%
%%%%
\begin{figure}[t]
\centering
\includegraphics[width=0.4\textwidth]{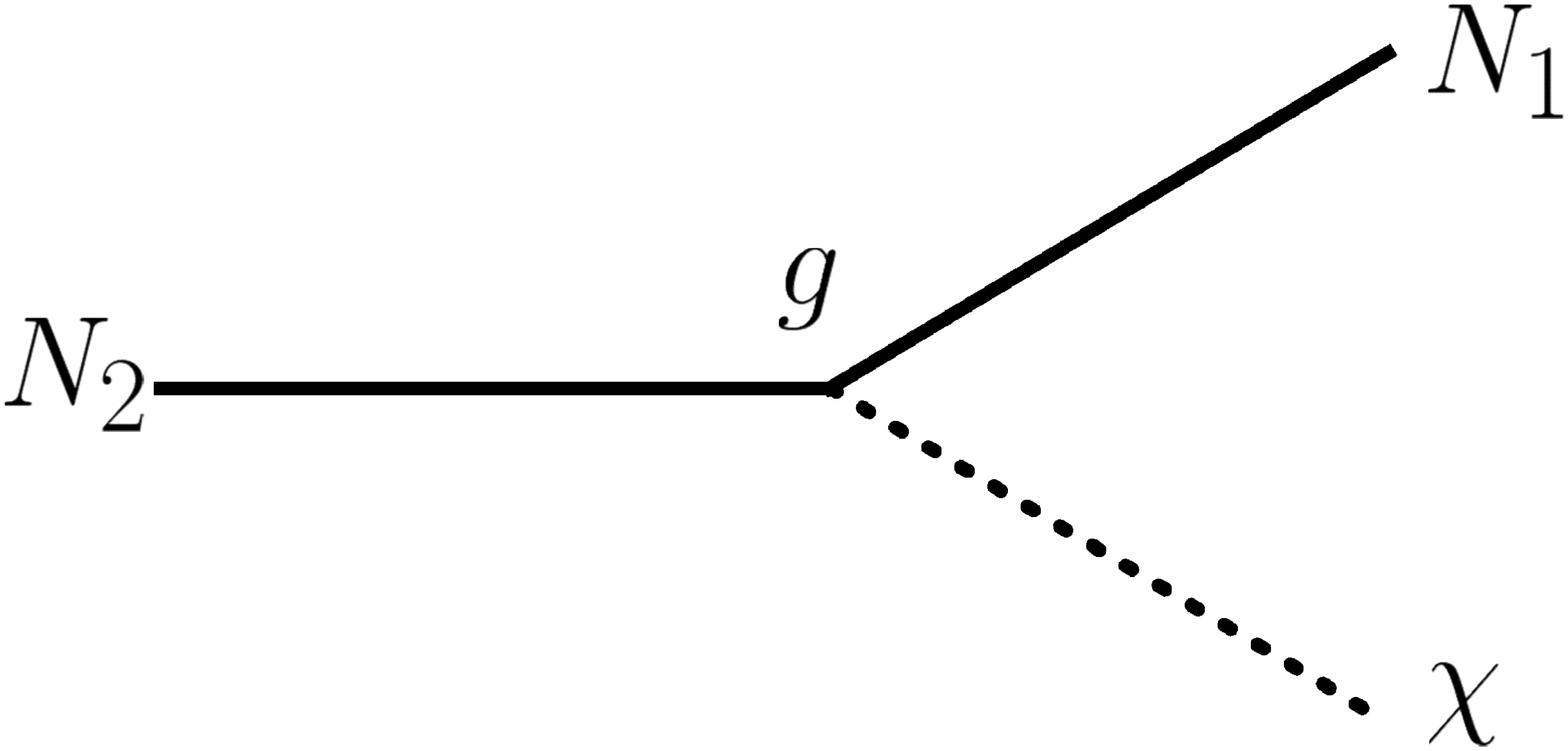}
\caption{
The Feynman diagram for Majorogenesis in the scenario (A).
}
\label{172708_30Mar20}
\end{figure}
%%%%
%

%
The DM creation process is $ N_{2} \overset{g}{\to} N_{1}  \chi ~(M_{N_{2}} > M_{N_{1}} + m_{\chi})$, 
which is shown in Fig.~\ref{172708_30Mar20}, 
and the decay width is given by
% -----
\begin{align}
 \Gamma_{N_{2} \to N_{1} \chi} = \frac{g^2 M_{N_2}}{32 \pi} I\biggl( \frac{ M_{N_{1}} }{ M_{N_{2}} }, \frac{ m_{\chi} }{ M_{N_{2}} } \biggr),
\end{align}
% -----
where the function $I(x,y)$ is defined by
$I (x, y)\equiv \left[ (1-x)^2 -y^2 \right]^{3/2} \left[ ( 1+ x)^2 -y^2 \right]^{1/2}$.
On the other hand,
the thermal creation process of the RH neutrinos are given by
$ N_{i} \longleftrightarrow L_{\alpha} H~(L_{\alpha}^{c} H^{\dagger})$,
and the decay width is expressed as
% -----
\begin{align}
 \Gamma_{N_i \to \mathrm{B}} = \frac{ |Y^{\nu}_{\alpha i}|^2 M_{N_i} }{8 \pi}.
 \label{171144_26Mar20}
\end{align}
% -----
%

%
The Boltzmann equations for the RH neutrinos and the Majoron are given by
% -----
\begin{align}
 \frac{d Y_{N_{2}}(x)}{dx} =& -\frac{ \Gamma_{N_{2} \to N_{1} \chi} }{ H x } \frac{K_{1}(r_{2} x)}{K_{2} (r_2 x)} Y_{N_{2}}(x)
 \nonumber\label{155305_26Mar20}\\
 &- \frac{ \Gamma_{N_{2} \to \mathrm{B}} }{ H x } \frac{K_{1} (r_2x)}{K_{2}(r_2x)} \bigl[ Y_{N_{2}}(x) - Y^{\mathrm{eq}}_{N_{2}}(r_2 x) \bigr],
 \\
 \frac{ d Y_{N_{1}}(x)}{dx} = & + \frac{ \Gamma_{N_{2} \to N_{1} \chi} }{H x} \frac{K_{1} (r_2 x)}{K_{2} (r_2 x)} Y_{N_{2}}(x)
 \nonumber\\
 &- \frac{\Gamma_{N_{1} \to \mathrm{B} } }{H x} \frac{K_{1} (r_1 x)}{ K_{2} (r_1 x)} \bigl[ Y_{N_{1}}(x) - Y^{\mathrm{eq}}_{N_{1}}(r_1 x) \bigr],
 \\
 \frac{ d Y_\chi (x)}{d x} =& + \frac{\Gamma_{N_{2} \to N_{1} \chi}}{H x} \frac{K_{1} (r_2 x)}{ K_{2} (r_2 x)} Y_{N_{2}}(x),\label{155324_26Mar20}
\end{align}
% -----
where $H$ denotes the Hubble parameter, $K_{n}$ is the modified Bessel function of the second kind.
We introduce the dimensionless parameter $x$ by $x \equiv m_{\chi} /T$ for the temperature $T$ and 
the mass ratios by $r_i \equiv M_{N_{i}} /m_{\chi}$.
The yield of a particle $X$ is defined by $Y_{X}\equiv n_X / s$ with $n_X$ and $s$ being the number density of $X$ and the entropy density, respectively.
The temperature dependence of the Hubble parameter and the entropy density is $ H = \sqrt{ 4\pi^3 g_{*} /45} T^2 / m_{\mathrm{Pl}}$ and $ s= 2\pi^2 g_{*}^{S} T^3/ 45$ with the Planck mass $m_{\mathrm{Pl}}$.
The function form of $Y_{X}$ in the thermal equilibrium is given by
% -----
\begin{align}
 Y^{\mathrm{eq}}_{X}(z) = g_{X} \biggl( \frac{45}{4 \pi^4 g_{*}^S} \biggr) z^2 K_{2}(z),
\end{align}
% -----
with $g_{X}$ being the number of the degrees of freedom for the particle $X$.
We assume that the SM particles are always in the thermal bath
and neglect the inverse decay $N_{1}  \chi \to N_{2}$ because the contribution from this process is small.
Using Eqs.~\eqref{155305_26Mar20}-\eqref{155324_26Mar20}, we obtain
% -----
\begin{align}
 Y_{\chi} (\infty) =& \int_{x_I}^{\infty} dx \frac{d Y_{\chi}(x)}{dx}
 \nonumber\\
 =& \frac{ \Gamma_{N_{2} \to N_{1} \chi} \Gamma_{N_{2} \to \mathrm{B}} }{ \Gamma_{N_2 \to N_1 \chi} +\Gamma_{N_2 \to \mathrm{B}} } \int_{x_I}^{\infty} dx\, \frac{1}{H x} \frac{K_{1} (r_2 x)}{K_{2} (r_2 x)} Y^{\mathrm{eq}}_{N_2}( r_2 x) 
 \label{eq:intYchi},
\end{align}
% -----
where we have assumed $Y_{N_{2}}(x_{I}) = Y_{N_{2}} (\infty) =0$.
The integral Eq.~(\ref{eq:intYchi}) can be carried out approximately and the Majoron relic abundance is evaluated as
% -----
\begin{align}
 Y_{\chi}(\infty) &\approx \biggl(\frac{ \Gamma_{N_{2} \to N_{1} \chi} \Gamma_{N_{2} \to \mathrm{B}} }{ \Gamma_{N_2 \to N_1 \chi} +\Gamma_{N_2 \to \mathrm{B}} } \biggr) \frac{ 405 \sqrt{5} }{ 8 \pi^{9/2} g_*^S g_*^{1/2}} \frac{m_{\mathrm{Pl}}}{M_{N_2}^2} \\
& \approx   \Gamma_{N_{2} \to N_{1} \chi} \frac{ 405 \sqrt{5} }{ 8 \pi^{9/2} g_*^S g_*^{1/2}} \frac{m_{\mathrm{Pl}}}{M_{N_2}^2}\label{173903_26Mar20} ,
\end{align}
% -----
where we have used $\Gamma_{N_{2} \to N_{1}\chi} \ll  \Gamma_{N_{2} \to \mathrm{B}} $ 
and the explicit expressions of the Hubble parameter and the entropy density.
%

%
%%%%%
\begin{figure}[t]
\centering
\includegraphics[width=0.48\textwidth]{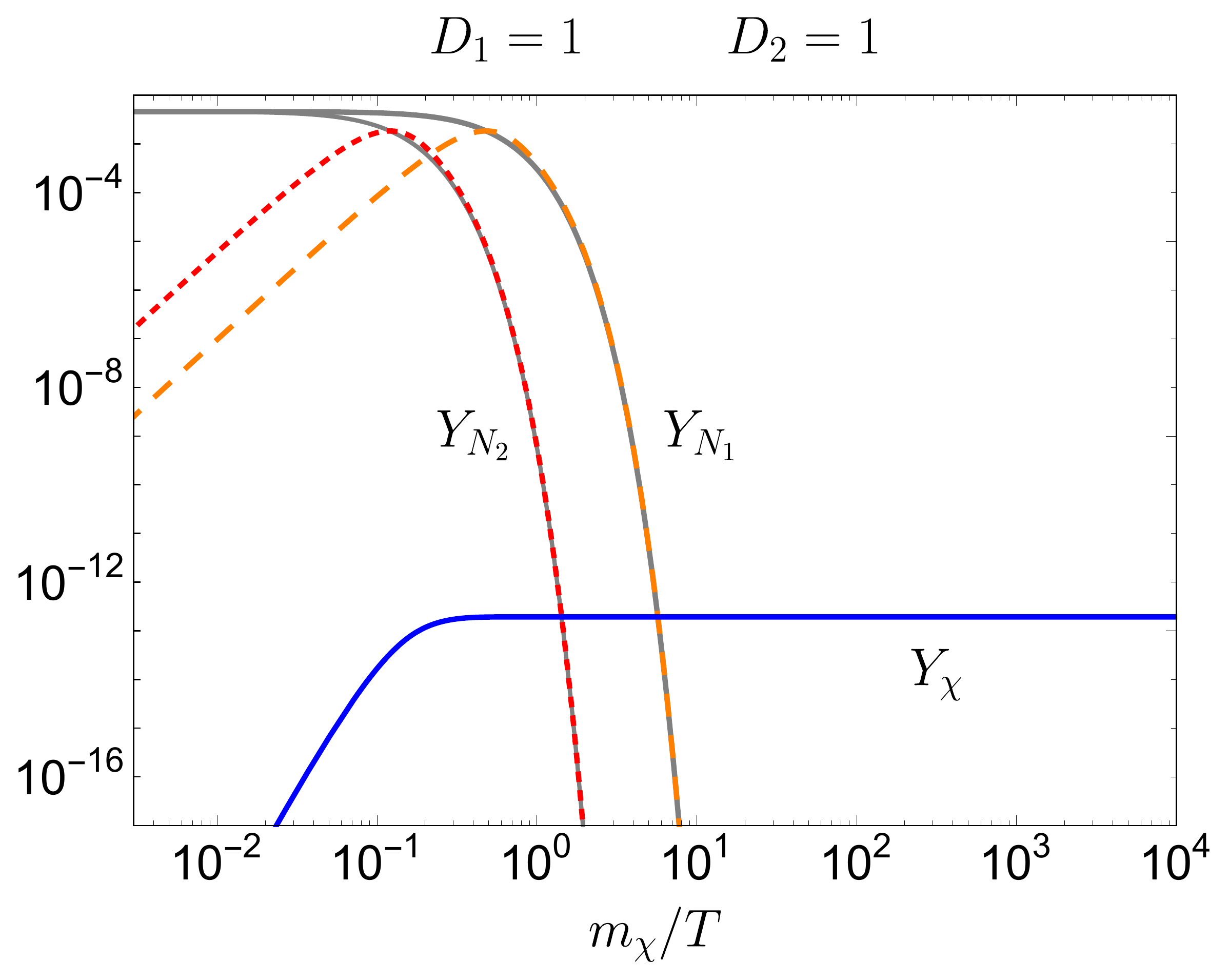}\hspace*{5mm}
\includegraphics[width=0.48\textwidth]{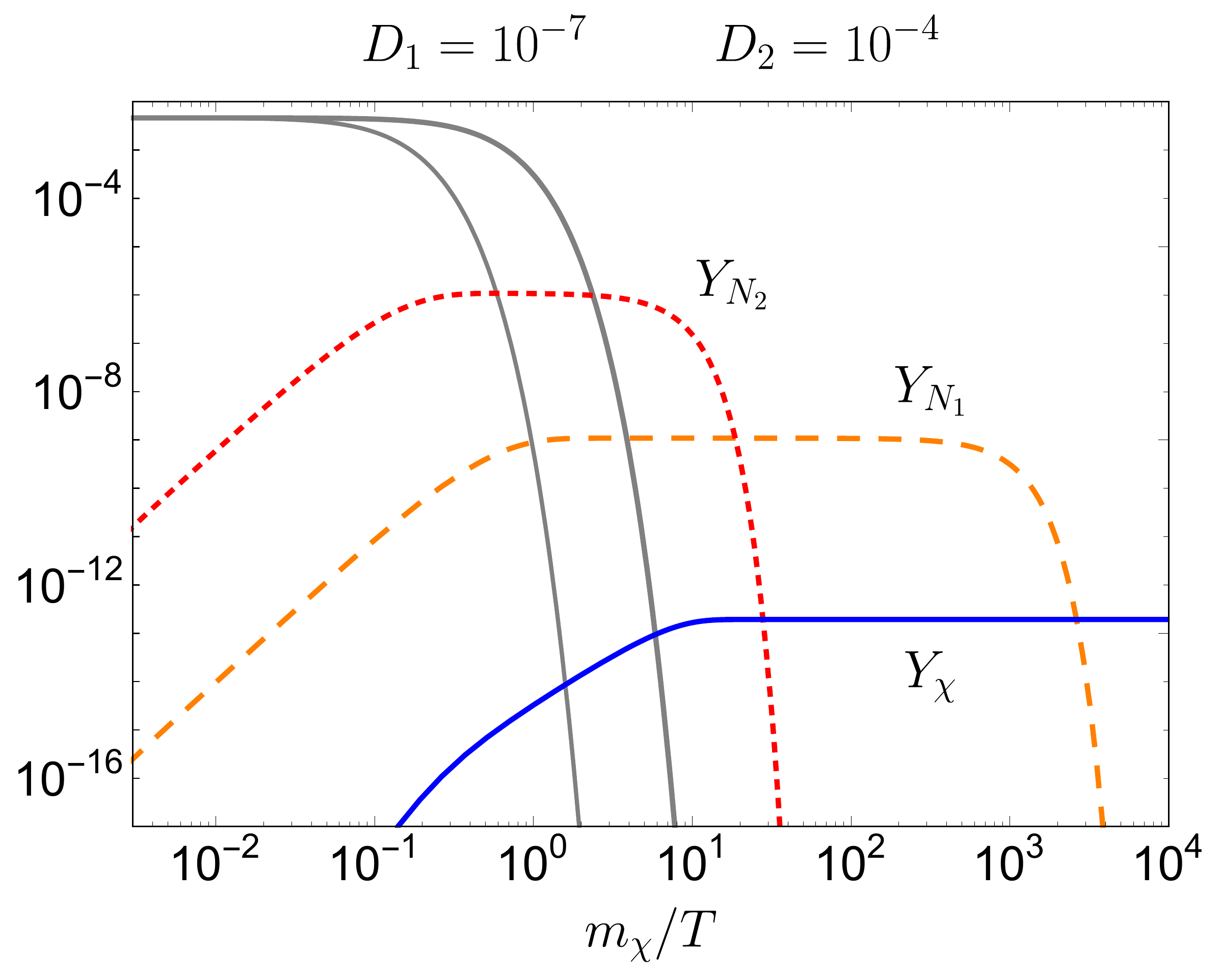}
\caption{%-----
The solutions of the Boltzmann equations with the masses $M_{N_1} =5\ \mathrm{TeV}$, $M_{N_2} = 20 \ \mathrm{TeV}$, $m_{\chi} =1 \ \mathrm{TeV}$ and the Yukawa coupling $g=10^{-11}$.
The black solid lines indicate the yields in the thermal equilibrium, $Y_{N_i}^{\mathrm{eq}}$.
The decay parameters $D_{i} = \Gamma_{N_{i} \to \mathrm{B}}/ H(T= M_{N_i})$ are changed.
In the left panel, the decay parameters are unity.
In the right panel, the decay parameters are hierarchical and tiny values.
}%-----
\label{fig:NNchi}
\end{figure}
%%%%%
%

%
The time evolution of the yields are shown in Fig.~\ref{fig:NNchi}.
The masses for the particle contents are fixed as $M_{N_{1}} = 5\ \mathrm{TeV}$, $M_{N_{2}} = 20 \ \mathrm{TeV}$ and $m_{\chi} =1\ \mathrm{TeV}$, and the off-diagonal Yukawa coupling is chosen as $g = 10^{-11}$.
The decay parameter $D_{i}$ is defined by the ratio of the decay width of $N_{i} \to \mathrm{SM}$ 
to the Hubble parameter $H$ as $D_{i} =\Gamma_{N_i \to \mathrm{B}}/ H ( T = M_{N_{i}})$ 
and is related to the neutrino Yukawa couplings $Y^{\nu}_{\alpha i}$ (see Eq.~\eqref{171144_26Mar20}).
In the left panel, the two decay parameters are unity, 
and then the RH neutrinos go into the thermal bath 
and the yields follow the thermal equilibrium distribution (black solid lines in the figure).
In the right panel, the two decay parameters are too small to put $Y_{N_i}$ into the thermal bath.
It is interesting that the final result $Y_{\chi}(\infty)$ converges to the same value 
independently of the magnitudes of the neutrino Yukawa couplings $Y^{\nu}_{\alpha i}$,
which is clear from Eq.~\eqref{173903_26Mar20}.
This is because, 
for small $Y^{\nu}_{\alpha i}$, the thermally induced amount of the RH neutrinos around their mass scale becomes small 
while the branching ratio decaying into the Majoron becomes large and these two effects are canceled out.
In the thermal historical point of view, 
the independence of neutrino Yukawa couplings is understood by the fact 
that thermally induced $N_{2}$ is proportional to $D_{2}$ and the time interval 
where the decay $N_{2} \to N_{1} \chi$ is effective is inversely proportional to $D_{2}$.
Then the relic abundance of the Majoron is given by
% -----
\begin{align}
 \Omega_{\chi} h^2 =& \frac{m_{\chi} Y_{\chi} (\infty) s_0}{\varepsilon_{c,0}/h^2} 
 \nonumber\\
 \approx&
 0.1075\times
 \biggl(\frac{g}{10^{-11}}\biggr)^2
 \biggl(  \frac{100}{g_*^S} \biggr)
 \biggl( \frac{100}{g_*} \biggr)^{1/2}
 \biggl(\frac{m_\chi}{1 \ \mathrm{TeV}}\biggr)
 \biggl(\frac{20 \ \mathrm{TeV}}{M_{N_2}}\biggr)
 I\biggl( \frac{M_{N_1}}{M_{N_2}}, \frac{m_\chi}{M_{N_2}} \biggr),
\end{align}
% -----
where $s_{0} = 2891 \ \mathrm{cm}^{-3}$ is the today entropy density, and $\varepsilon_{c, 0}=5.16 ( h / 0.7 )^2 \ \mathrm{GeV}\ \mathrm{m}^{-3}$ is the today critical energy density.
The current observed value of DM abundance is $\Omega_{\mathrm{DM}} h^2 = 0.1200(12)$ \cite{Aghanim:2018eyx}.
As we stated above, the relic abundance is independent of the neutrino Yukawa couplings $Y^{\nu}_{\alpha i}$.
%

%
%%%%%%
\begin{figure}
\centering
\includegraphics[width=0.45\textwidth]{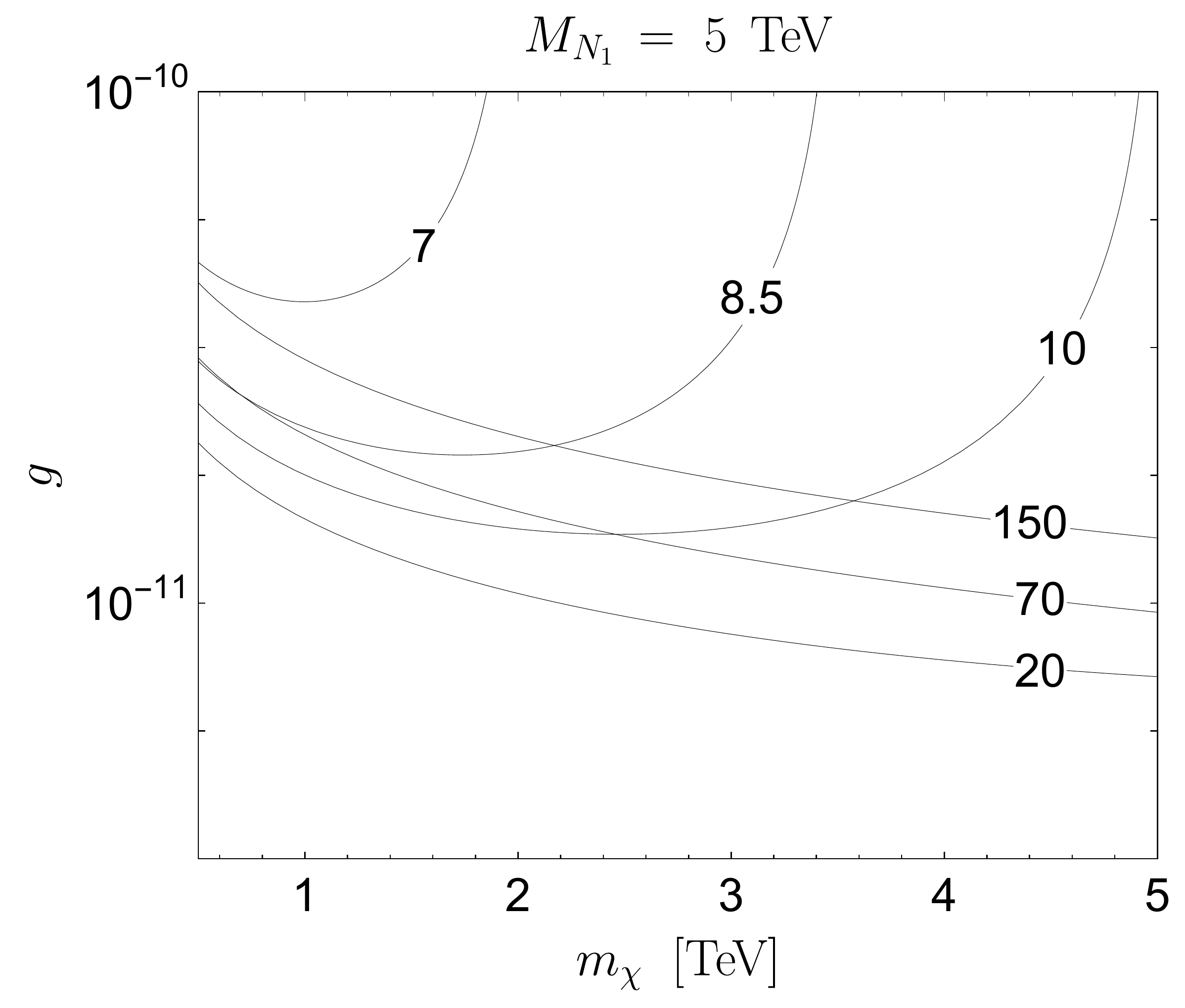}
\hspace{1em}
\includegraphics[width=0.43\textwidth]{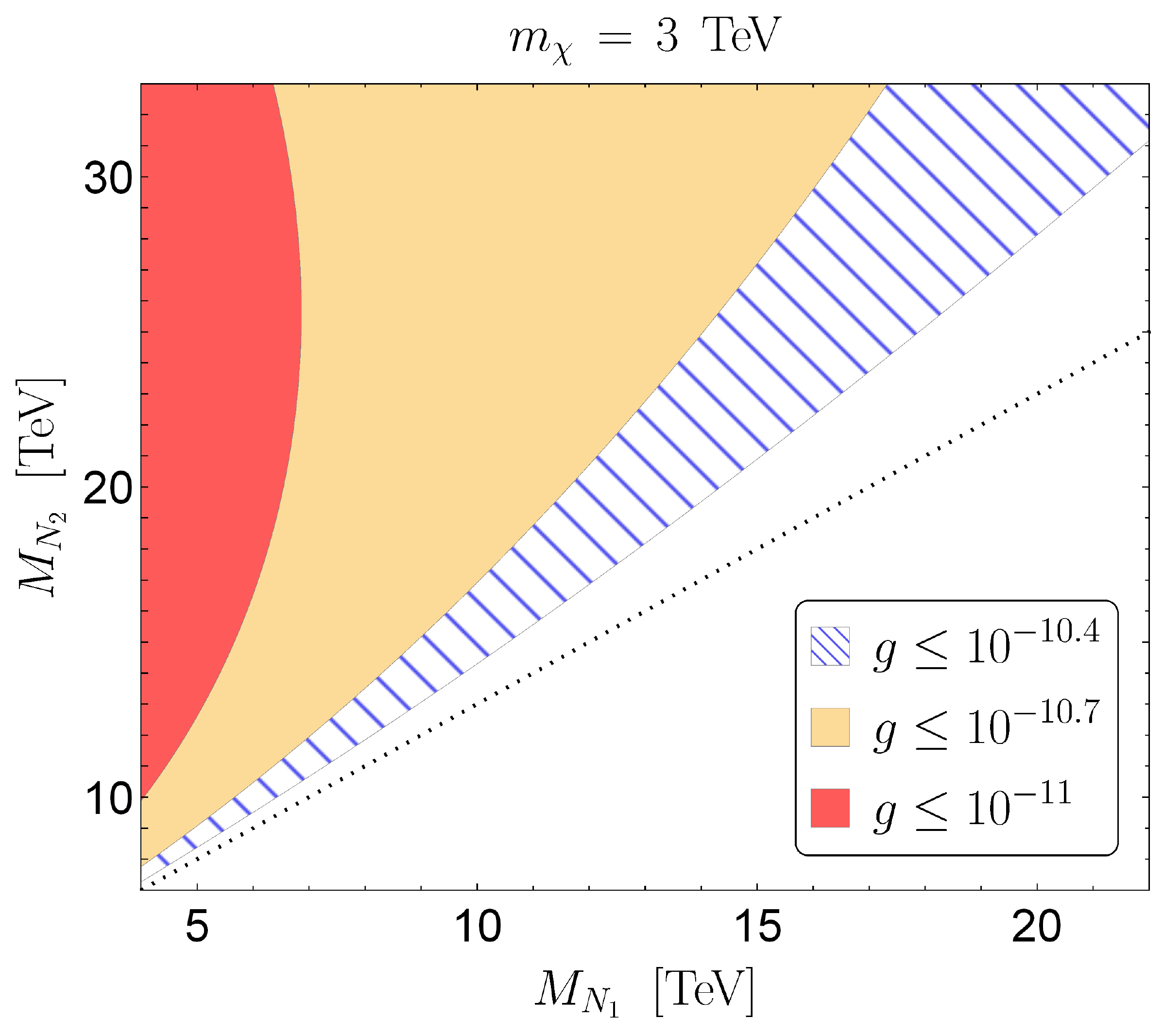}
\caption{% -----
The allowed region to realize the DM relic abundance in $(m_{\chi}, g)$- and $(M_{N_1}, M_{N_2})$-planes.
See the text for details.
}% -----
\label{fig:fMMN}
\end{figure}
%%%%%%
%

%
In Fig.~\ref{fig:fMMN}, we show the allowed parameter regions in $(m_{\chi}, g)$ and $(M_{N_{1}}, M_{N_{2}} )$ planes.
In the left panel, each line represents the parameter space realizing the DM relic abundance
for several choices of $M_{N_2}$ with $M_{N_{1}}$ fixed as $5\ \mathrm{TeV}$.
Note that the stronger Yukawa coupling $g$ is required for the smaller Majorana mass $M_{N_2}$
since the phase factor $I(M_{N_1}/M_{N_2},m_\chi/M_{N_2})$ becomes smaller.
On the other hand, large $g$ is also required for larger $M_{N_2}$
because $I \sim 1$ and the relic abundance is inversely proportional to $M_{N_2}$.
The allowed region regarding $M_{N_2}$ as a free parameter is bounded from below
by the critical line corresponding to $M_{N_{2}} \sim 20 \ \mathrm{TeV}$.
Thus the lower bound for $g$ is around $g \sim 10^{-11}$.
In the right panel, we show the allowed region with the DM mass $m_{\chi} = 3\ \mathrm{TeV}$ 
and the Yukawa coupling $ g \leq 10^{-10.4},\  10^{-10.7}, \  10^{-11}$.
If we take a severer bound for the off-diagonal Yukawa coupling $g \leq 10^{-11}$
(red region in the figure), 
the lightest RH neutrino mass has to be in $3 \ \mathrm{TeV} \leq M_{N_1} \leq 7 \ \mathrm{TeV}$, 
and the mass $M_{N_2}$ has to be larger than 10 TeV\@.
Interestingly, the bound on $g$ for this scenario to work is marginally comparable 
with the experimentally constrained upper bound Eq.~\eqref{202732_26Mar20}.
Therefore, the scenario (A) can be proved or excluded in the near future observations.
%

%%%
\subsection{(B): Scalar portal interaction}
\label{subsec:scalar}

Let us move to another scenario, in which we introduce the mixing coupling $\lambda_{H\Phi} |H|^2 |\Phi|^2$.
We consider the freeze-in creation of the Majoron in this model.
The scalar potential is written as
% -----
\begin{align}
 V(H, \Phi) =& V_{H}(H)-\frac{\mu_{\Phi}^2}{2} |\Phi|^2 + \frac{\lambda_{\Phi}}{2} |\Phi|^4
 + \lambda_{H\Phi} |H|^2 |\Phi|^2
 -\frac{m^2}{4} \bigl( \Phi^2 + {\Phi^*}^2 \bigr),
\end{align}
% -----
and the conditions for the quartic couplings such that the potential is bounded from below are
$ \lambda_{H} >0,~ \lambda_{\Phi} >0, ~ \sqrt{\lambda_{H} \lambda_{\Phi}} + \lambda_{H\Phi} >0$.
In addition, the quartic coupling $\lambda_{\Phi}$ has the upper bound $8 \pi /3$ 
from the perturbative unitarity as discussed in Ref.~\cite{Chen:2014ask}.
The quartic coupling $\lambda_{H\Phi}$ is also constrained by the bound of the mixing angle between the CP-even components \cite{Falkowski:2015iwa}.
%

%
%%%%
\begin{figure}[t]
\centering
\includegraphics[width=0.4\textwidth]{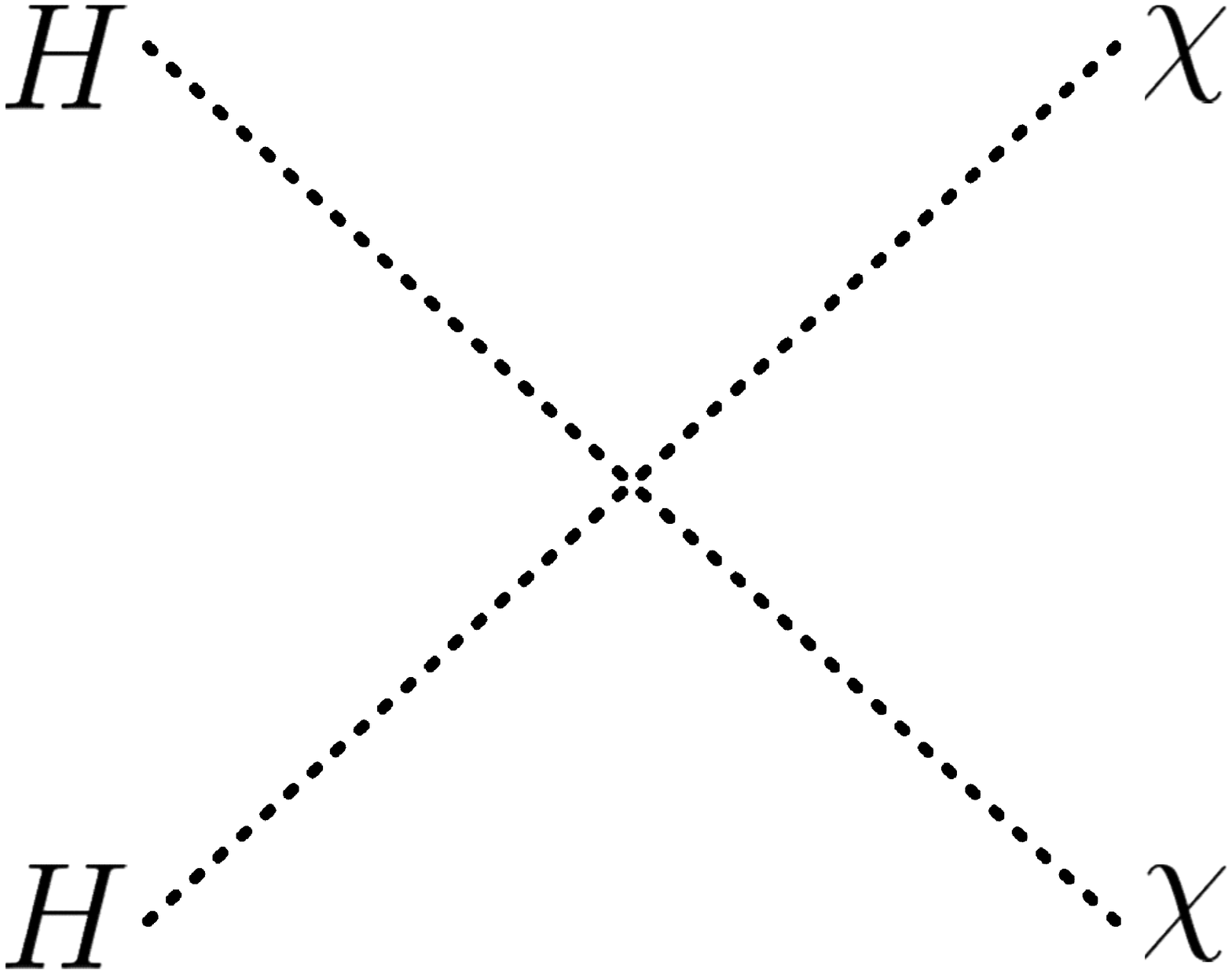}\hspace{2em}
\includegraphics[width=0.4\textwidth]{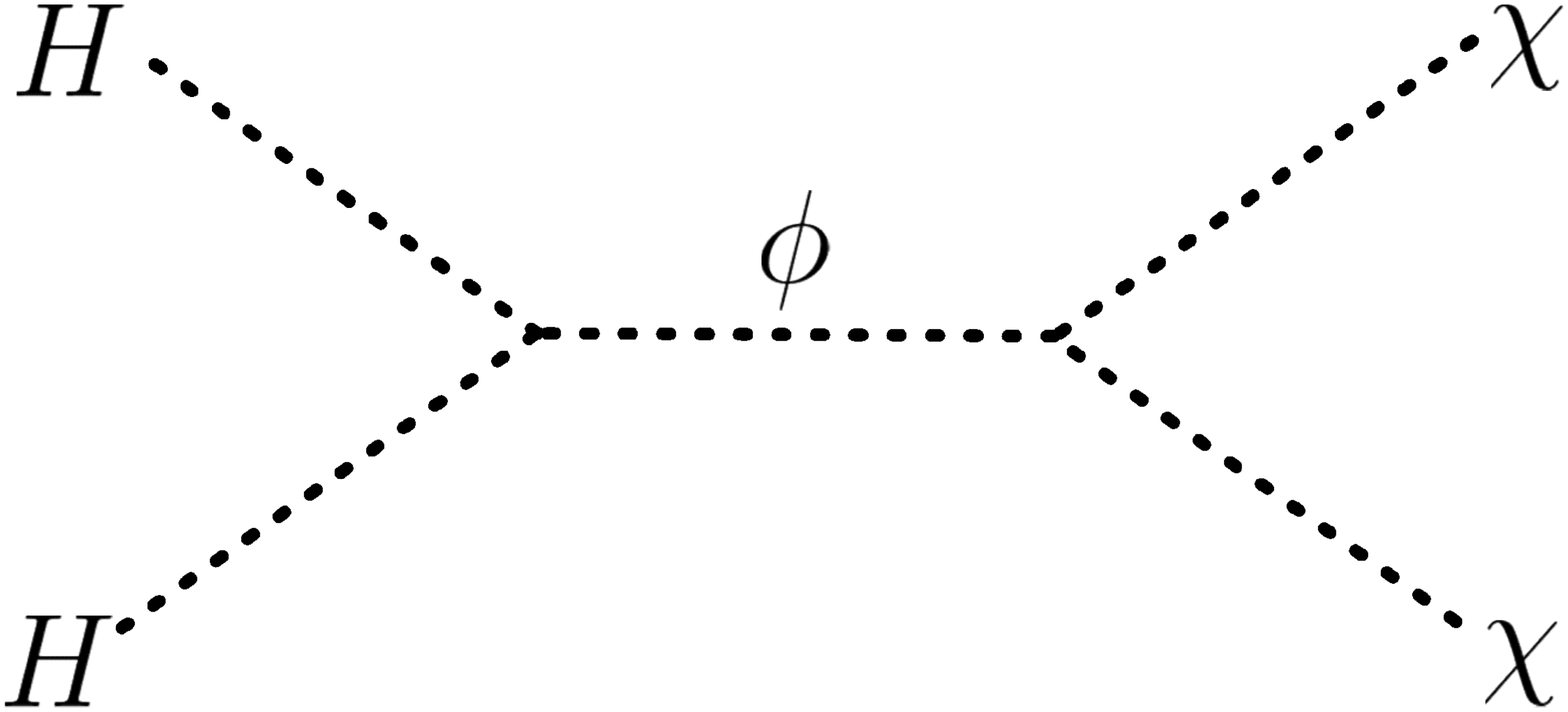}
\caption{
The Feynman diagrams for the Majorogenesis in the scenario (B).
The contact type interaction in the left panel is canceled by the low-energy contribution from the scalar mediated interaction in the right panel.
}
\label{fig:NGBint}
\end{figure}
%%%%
%

%
One important feature of the Majoron is a cancellation due to the nature of NGB
in two-body scattering processes such as Fig.~\ref{fig:NGBint}.
The contribution from the contact type four-point interaction (left panel) is canceled 
by the one from the $\phi$-mediated interaction (right panel) in the soft limit, 
and the remaining value is suppressed by the large decay constant.
Indeed, the leading contribution after the cancellation comes from the portal energy in the propagator, which is written as
% -----
\begin{align}
 i \mathcal{M} ( H^\dagger H \to \chi \chi )(s)  =  -i\frac{\lambda_{H\Phi}}{s-m_{\phi}^2} s ,
 \label{eq:NGBamp}
\end{align}
% -----
where $s$ is the Mandelstam's $s$ variable and $m_{\phi}^2 = \lambda_{\Phi} v_{\phi}^2$ is the mass of $\phi$.
This is consistent with the result implied by the soft-pion theorem,
and is easily understood in the non-linear representation:
% -----
\begin{align}
 \Phi = \frac{ v_{\phi} + \phi }{\sqrt{2}} e^{i \pi /v_{\phi}}.
\end{align}
% -----
The phase field $\pi$ is the Majoron in this representation and is the same as $\chi$ to the leading order of $1/ v_{\phi}$.
We have the following interaction vertices in the Lagrangian:
% -----
\begin{align}
 \mathcal{L}_{\mathrm{int}} \supset  \frac{\phi}{v_{\phi}} \Bigl[ (\partial_\mu \pi)^2 -m_{\chi}^2 \pi^2 \Bigr] 
 - \lambda_{H\Phi} v_{\phi} \phi |H|^2,\label{201509_27Mar20}
\end{align}
where the derivative coupling between the (p)NGB $\pi$ and the CP-even scalar particle $\phi$
has come from the kinetic term of $\Phi$.
% -----
The scattering amplitude for $H H \to \pi \pi$ evaluated from this interaction Lagrangian  Eq.~\eqref{201509_27Mar20} is the same as Eq.~(\ref{eq:NGBamp}), 
which is now given by a single diagram like the right-panel of Fig.~\ref{fig:NGBint} and the energy ($s$) dependence originates from the derivative coupling.
The Boltzmann equation for the Majoron DM is given by
% -----
\begin{align}
 \frac{d Y_{\chi}(x)}{dx} =& \frac{2}{s H x} \gamma^{\chi \chi}{}_{H^\dagger H},
\end{align}
% -----
where $\gamma^{\chi \chi}{}_{H^\dagger H}$ is the interaction density defined as
% -----
\begin{align}
 \gamma^{\chi\chi}{}_{H^\dagger H} \equiv &
 \frac{4}{2! 2!}\frac{T \lambda_{H\Phi}^2}{2^{9} \pi^5} \int_{4 m_{\chi}^2}^{\infty} ds\,
 \sqrt{s- 4m_{\chi}^2} K_{1}(\sqrt{s}/T)
 \frac{s^2}{(s-m_{\phi}^2)^2}.
\end{align}
% -----
The prefactor 4 comes from the degrees of freedom of the Higgs doublet in the symmetric phase.
We integrate the Boltzmann equation with the initial condition $Y_{\chi} (x_{R})=0$, then the Majoron abundance can be analytically evaluated as
% -----
\begin{align}
 \Omega_{\chi} h^2 
 \approx & 1.5 \times 10^{25}\ \mathrm{GeV}\ 
 \biggl( \frac{100}{g^{S}_{*}} \biggr)
 \biggl( \frac{100}{g_*} \biggr)^{1/2}
 \biggl( \frac{m_{\chi}}{1 \ \mathrm{TeV}} \biggr)
 \frac{\lambda_{H\Phi}^2 T_{R}^3}{m_{\phi}^4}
 ,\label{220156_26Mar20}
\end{align}
% -----
where $T_{R}$ is the reheating temperature satisfying $x_R \equiv m_{\chi} /T_{R}$.
In Eq.~\eqref{220156_26Mar20}, we have assumed that $m_\phi$ is larger than the reheating temperature.
Due to the energy dependence of the amplitude (\ref{eq:NGBamp}) and the heavy portal scalar, 
the relic abundance is dominated by the contribution from the ultraviolet(UV)-region unlike the previous case (A)
and depends on the reheating temperature $T_{R}$, which arises from the UV physics.
A similar type of freeze-in effect is discussed in the context of higher dimensional operators \cite{Hall:2009bx}.
%

%
%%%
\begin{figure}[t]
\centering
\includegraphics[width=0.55\textwidth]{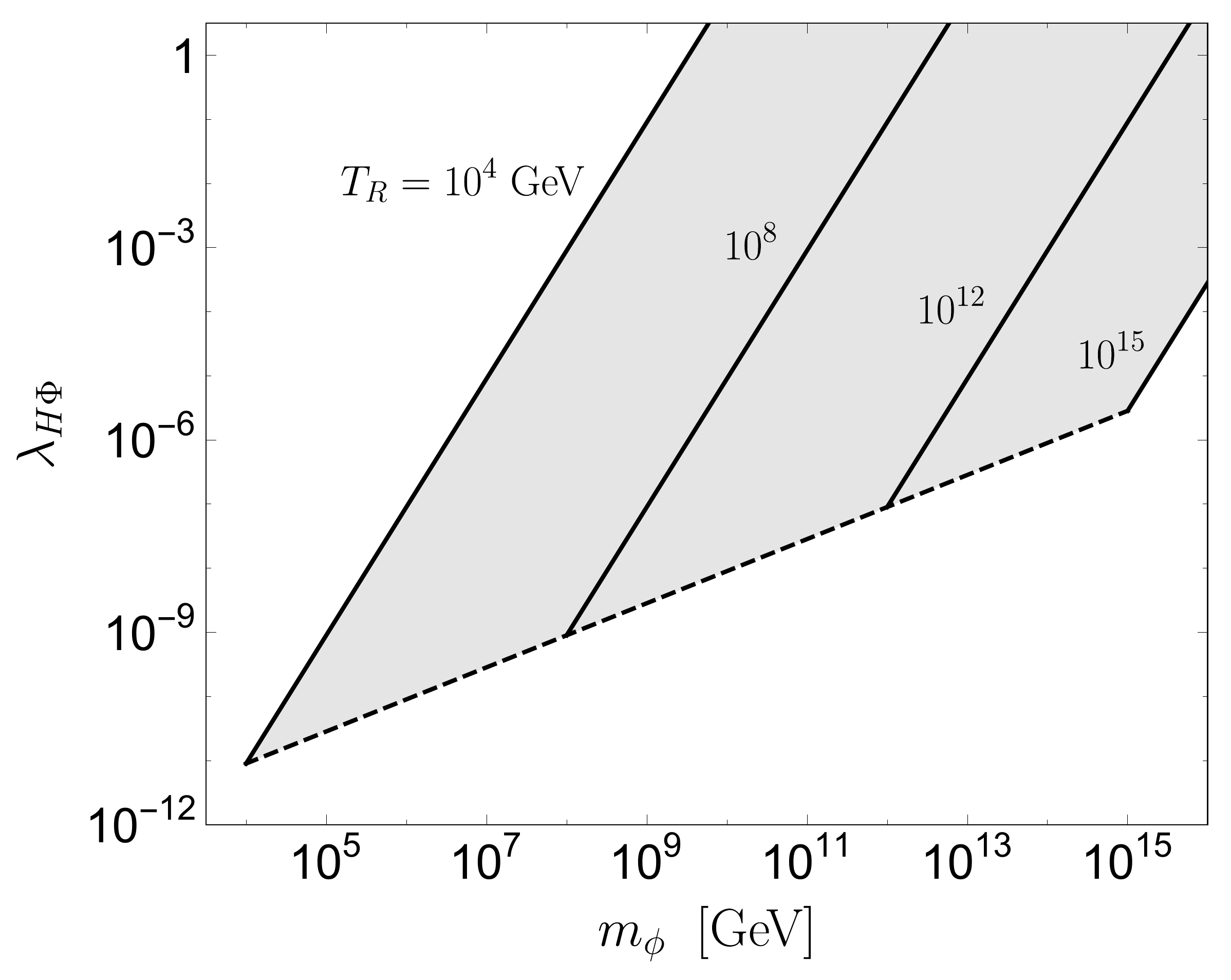}
\caption{% -----
The parameter space in the $(m_{\phi}, \lambda_{H\Phi})$ plane explaining the DM relic by the TeV scale Majoron.
See the text for details.
}% -----
\label{fig:lambdaHPhi}
\end{figure}

In Fig.~\ref{fig:lambdaHPhi}, we show the allowed region in the $( m_{\phi}, \lambda_{H\Phi})$ plane.
Each solid line represents the parameter space realizing the DM relic abundance 
and the region above each line for $T_{R}$ being fixed is excluded by the over creation.
The dashed line means the case $m_{\phi} = T_{R}$, and the region below this line is not valid 
because we assumed that the mass of $\phi$ is larger than the reheating temperature such that $\phi$ is inactive in thermal evolution after the reheating.
The shaded region shows the parameter space in which the DM relic abundance is realized
regarding the reheating temperature as a free parameter.
The region of $m_\phi$ is taken as $10^{4}\ \mathrm{GeV} \leq m_{\phi} \leq 10^{15}\ \mathrm{GeV}$.
We note the VEV $v_{\phi}$ is large for the TeV-scale Majoron 
and the constraint from the mixing among the CP-even components is negligible due to the suppression by $v_{\phi}$.
We here give a comment on other previous work.
The portal-like coupling of pNGB has also been discussed in various contexts, e.g., Refs.~\cite{Gu:2010ys,Queiroz:2014yna}.
The existence of the contact type interaction by the quartic coupling $\lambda_{H\Phi}$ is usually assumed, 
but that is canceled by heavy scalar mediated contribution, as stated above.
Consequently, it seems that the thermal freeze-out creation and collider search of the pNGB DM are inaccessible in case of the large decay constant.
%

%%%
\subsection{(C): Resonant creation from non-thermal source}
\label{subsec:inf}

Let us consider the third scenario that the Majoron DM is created by the RH neutrino annihilation process
mediated by the heavy CP-even scalar $\phi$.
We here assume that the mass $m_\phi$ is smaller than the reheating temperature $T_R$ 
so that $\phi$ plays an important role in the thermal history of the universe. 
In this subsection, we consider the case of one generation RH neutrino for simplicity,
but the generalization to three generations RH neutrino is straightforward.
We further assume that the RH neutrino has a Yukawa coupling to the inflaton field $\varphi$ with mass $m_{\varphi}$.
This coupling generates the RH neutrinos non-thermally during the reheating,
and the yield at $T_{R}$ is evaluated as
% -----
\begin{align}
 Y_{N} = \frac{3}{2} \frac{T_{R}}{m_{\varphi}} \mathrm{Br}(\varphi \to NN).
\end{align}
% ----
Here $\mathrm{Br}(\varphi \to NN)$ is the branching ratio of $\varphi \to NN$ process,
which is given by $\Gamma_{\varphi \to NN}/ \Gamma_{\varphi}$
with $\Gamma_{\varphi}$ being the total decay width of the inflaton.
The reheating temperature $T_{R}$ is defined by $H( T= T_{R} ) = \Gamma_{\varphi}$.
%

%
%%%%
\begin{figure}[t]
\centering
\includegraphics[width=0.3\textwidth]{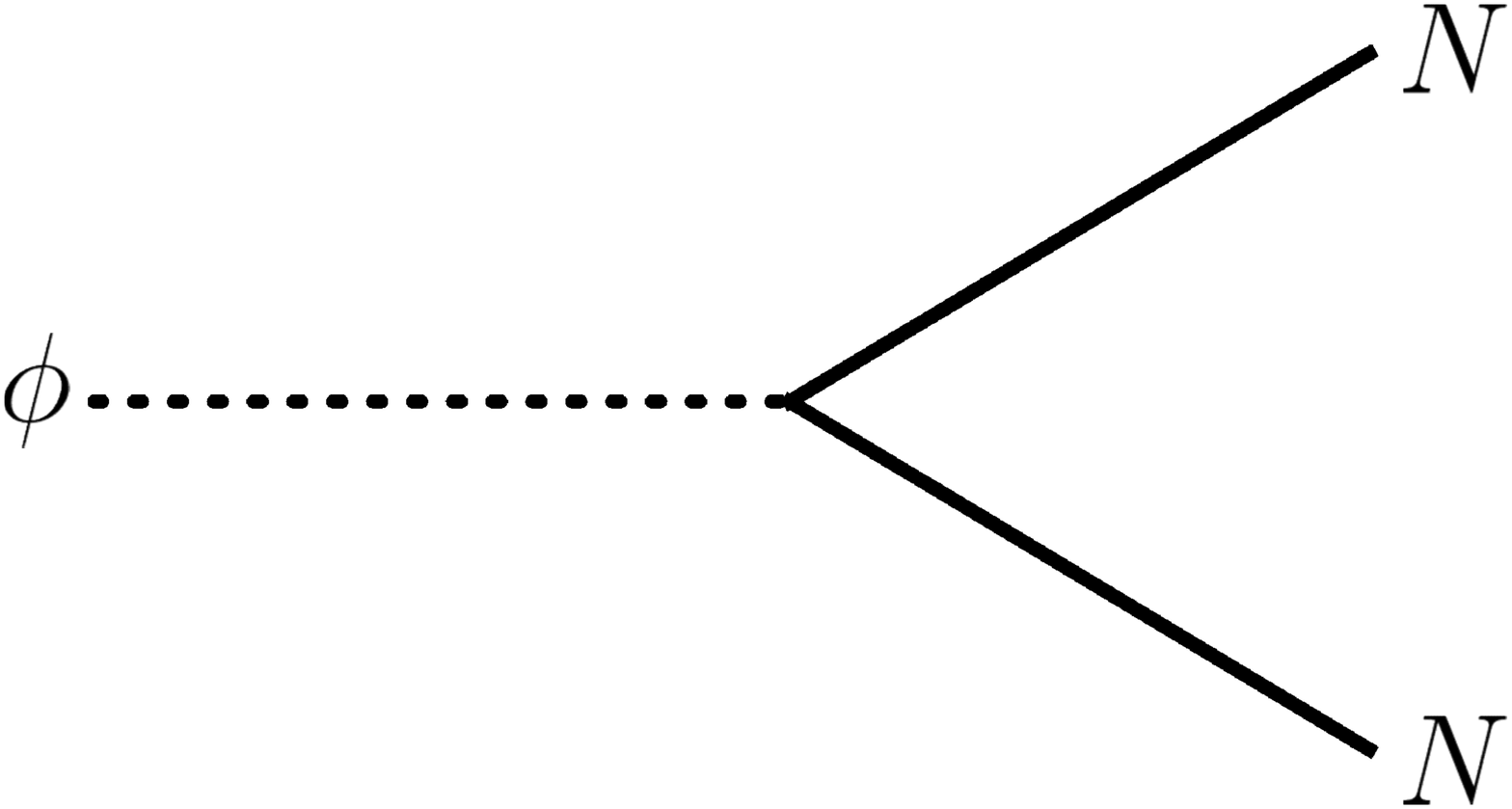}
\includegraphics[width=0.3\textwidth]{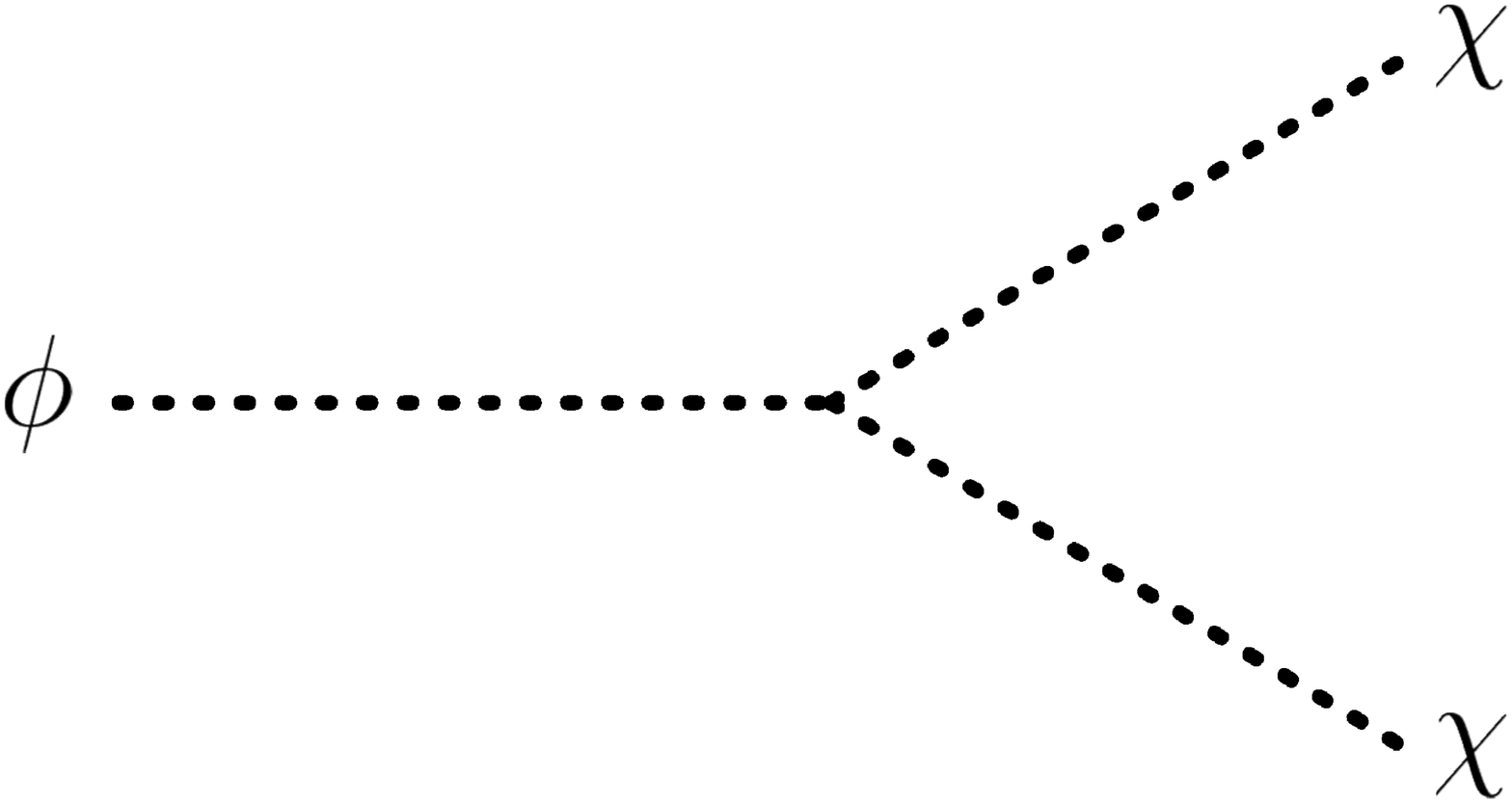}
\includegraphics[width=0.3\textwidth]{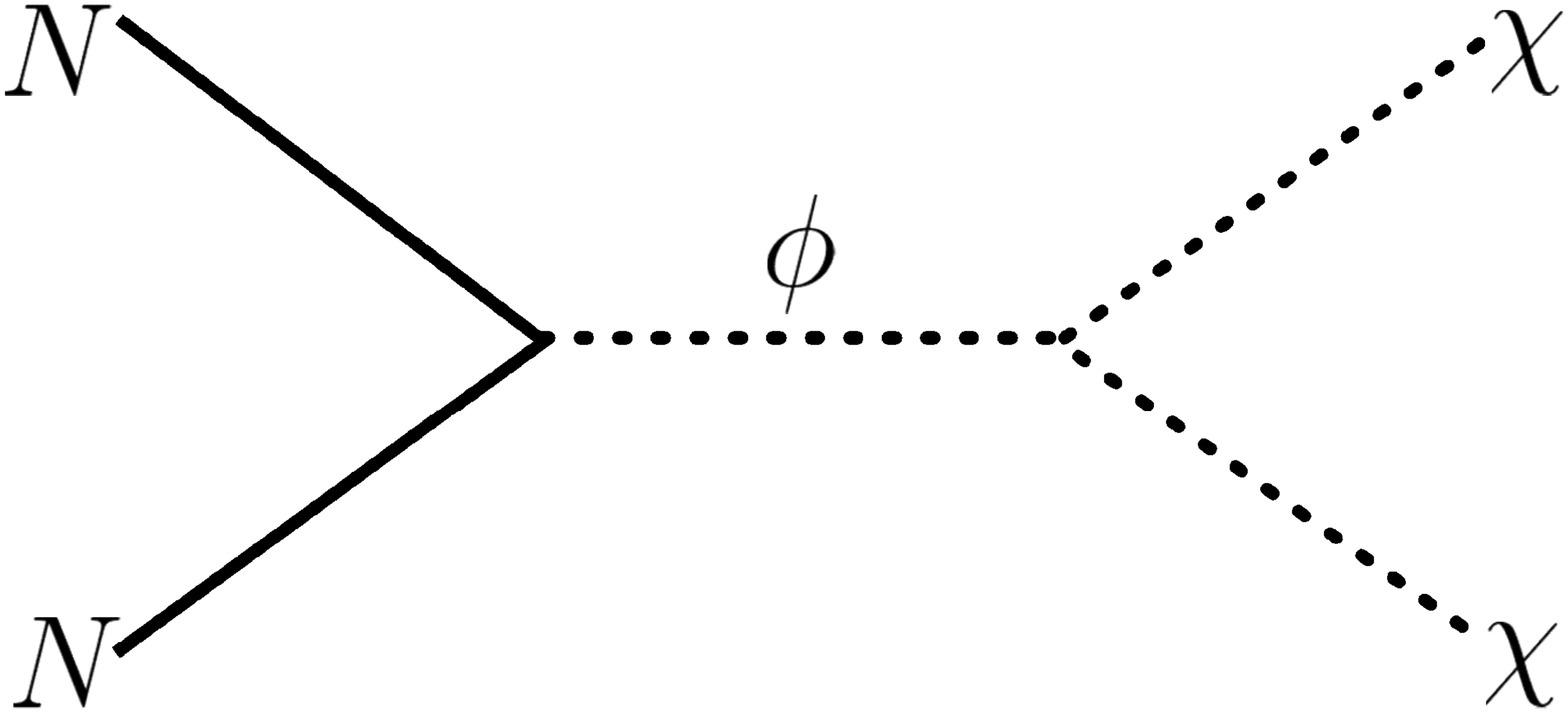}
\caption{% -----
The Feynman diagrams for the Majorogenesis in the scenario (C).
}
\label{192024_30Mar20}% -----
\end{figure}
%%%%
%

%
The RH neutrinos created by the inflaton can annihilate into the Majoron
through the scattering process mediated by $\phi$:
$ NN \overset{\phi}{\longleftrightarrow} \chi \chi$ 
as shown by Fig.~\ref{192024_30Mar20}.
The Yukawa coupling $f$ corresponding to $\phi \bar{N} N$ should be small from astrophysical constraints, 
and the three point coupling $\phi \chi \chi$ is also suppressed.
As we will see in the following, 
even for these tiny couplings, 
a sufficient amount of the Majoron DM can be generated with the resonant contribution of $\phi$.
The partial decay widths of $\phi$ to RH neutrinos and Majoron
% -----
\begin{align}
 \Gamma_{\phi \to N N } =
 \frac{f^2 m_{\phi}}{32 \pi} \biggl[ 1 - \frac{4 M_{N}^2}{m_{\phi}^2} \biggr]^{3/2},
 \quad
 \Gamma_{\phi \to \chi \chi} =&
 \frac{\lambda_{\Phi} m_{\phi}}{32 \pi} \biggl[1- \frac{4 m_{\chi}^2}{m_{\phi}^2} \biggr]^{1/2}.
\end{align}
% -----
The contribution to the Boltzmann equations from the $\phi$ portal annihilation process,
$N N \overset{\phi}{\to} \chi \chi$, is evaluated as
% -----
\begin{align}
 \gamma^{N N}{}_{\chi \chi} =&
 \frac{\lambda_{\Phi} f^2 T}{2^{11} \pi^5} \int_{4 M_{N}^2}^{\infty}
 ds\, \frac{ ( s- 4 M_{N}^2) ^{3/2} ( s- 4m_{\chi}^2)^{1/2}}{ s^{1/2} (s -m_{\phi}^2)^2}
 K_{1} (\sqrt{s} /T)
 \nonumber\\
 \approx &
 \frac{f^2 m_{\phi}^3 T}{2^6 \pi^3} \biggl[ 1- \frac{ 4 M_{N}^2 }{ m_{\phi}^2 } \biggr]^{3/2}
  K_{1} (m_{\phi} / T),
\end{align}
% -----
where we use the narrow width approximation.
%

%
%%%%%
\begin{figure}[t]
\centering
\includegraphics[width=0.5\textwidth]{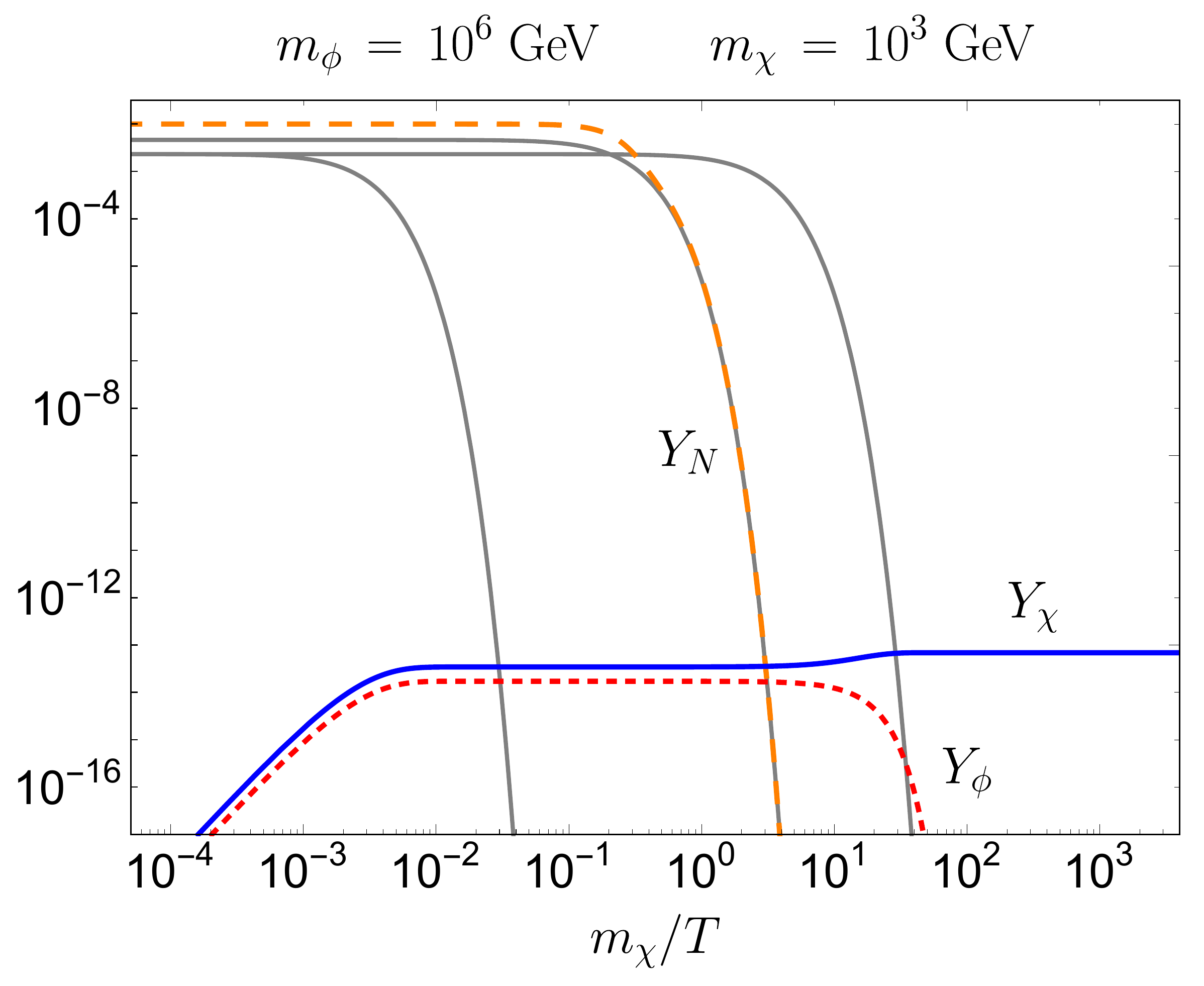}
\caption{%-----
The solution of the Boltzmann equations with $m_{\phi} = 10^{6}\ \mathrm{GeV}$, $m_{\chi} = 10^{3}~ \mathrm{GeV}$ and the initial value of the RH neutrino yield $Y_{N}(x_{R}) = 10^{-2}$.
The RH neutrino mass is fixed as $M_{N} = 10^{4} \ \mathrm{GeV}$, but the DM abundance $Y_{\chi}(\infty)$ does not depend on it.
}%-----
\label{fig:nonthermal}
\end{figure}
%%%%%
%

The Boltzmann equations for the Majorogenesis in this system are expressed as
% -----
\begin{align}
 \frac{d Y_{N}(x)}{d x} =&
 2 \frac{\Gamma_{\phi \to NN}}{Hx} \frac{ K_{1} (r_{\phi}x) }{K_{2} (r_{\phi} x)} Y^{\mathrm{eq}}_{\phi} (r_{\phi} x)
 \biggl[ \frac{ Y_{\phi} (x)}{Y_{\phi}^{\mathrm{eq}} (r_{\phi} x)}
 - \biggl( \frac{ Y_{N}(x) }{ Y_{N}^{\mathrm{eq}}(r x)} \biggr)^2 \biggr]
 \nonumber\\
 & - \frac{2}{H s x} \gamma^{NN}{}_{\chi\chi}
  \biggl( \frac{ Y_{N}(x) }{ Y_{N}^{\mathrm{eq}} (rx)} \biggr)^2
  - \frac{\Gamma_{N \to\mathrm{B}}}{Hx} \frac{ K_{1} (rx) }{K_{2} (rx) }
 \bigl[ Y_{N}(x) - Y_{N}^{\mathrm{eq}} ( r x) \bigr],
 \\
 \frac{ d Y_{\phi} (x) }{dx} =&
 - \frac{\Gamma_{\phi \to NN}}{Hx} \frac{K_{1} (r_{\phi} x)}{K_{2} (r_{\phi} x)} Y^{\mathrm{eq}}_{\phi} (r_{\phi} x)
 \biggl[ \frac{ Y_{\phi}(x) }{ Y^{\mathrm{eq}}_{\phi} (r_{\phi} x) }
 - \biggl( \frac{ Y_{N}(x) }{Y^{\mathrm{eq}}_{N} (rx)} \biggr)^2 \biggr]
 \nonumber\\
 &- \frac{ \Gamma_{\phi \to \chi \chi} }{Hx} \frac{K_{1}(r_{\phi} x)}{K_{2}(r_{\phi} x)}
 Y_{\phi}(x),
 \\
 \frac{d Y_{\chi}(x)}{dx} = & 2 \frac{\Gamma_{\phi \to \chi \chi}}{Hx} \frac{K_{1}(r_{\phi}x)}{K_{2}(r_{\phi}x)}
 Y_{\phi}(x)
  + \frac{2}{H s x} \gamma^{NN}{}_{\chi \chi}
 \biggl( \frac{Y_{N}(x)}{Y^{\mathrm{eq}}_{N} (rx)} \biggr)^2,
 \end{align}
% -----
where we have assumed that the SM particles are in the thermal bath.
$r_{\phi}$ and $r$ are defined as $r_{\phi} \equiv m_{\phi} / m_{\chi}$ and $r \equiv M_N / m_{\chi}$, respectively.
The relic density of the Majoron DM is found by solving these equations and evaluated approximately as
% -----
\begin{align}
 Y_{\chi} (\infty) \approx \frac{f^2 m_{\phi}}{16\pi} \biggl( 1 -\frac{4 M_{N}^2}{m_{\phi}^2} \biggr)^{3/2} \int_{x_R}^{\infty} dx\, \frac{2}{Hx} \frac{K_{1}(r_{\phi} x)}{K_{2}(r_{\phi}x)} Y^{\mathrm{eq}}_{\phi} (r_{\phi} x) \biggl( \frac{Y_{N}(x)}{Y^{\mathrm{eq}}_{N} (rx ) } \biggr)^2,
\end{align}
% -----
where we have used the boundary conditions $Y_{\phi}(x_{R} ) =Y_{\chi}(x_{R}) = Y_{N} (\infty) = Y_{\phi}(\infty)=0$.
The final result is given by
% -----
\begin{align}
 Y_{\chi}(\infty) \approx \frac{\pi^{5/2} g_{*}^S}{128 \sqrt{5} g_*^{1/2}} \frac{m_{\mathrm{Pl}}}{m_{\phi}} f^2 Y_{N}(x_R)^2 .
\end{align}
% -----
The relic abundance of the Majoron DM is evaluated as
% -----
\begin{align}
 \Omega_{\chi} h^2
 \approx & 4.02 \times 10^{27}\, \biggl( \frac{100}{g^S_*}\biggr) \biggl( \frac{100}{g_*} \biggr)^{1/2} \frac{m_{\chi}}{m_{\phi}} f^{2} Y_{N}(x_R)^2.
\end{align}
% -----
This result depends on the Yukawa coupling $f$, the scalar mass $m_{\phi}$, and the initial amount of RH neutrinos, but is independent of the RH neutrino mass.
%

%
%%%%%
\begin{figure}[t]
\centering
\includegraphics[scale=0.35]{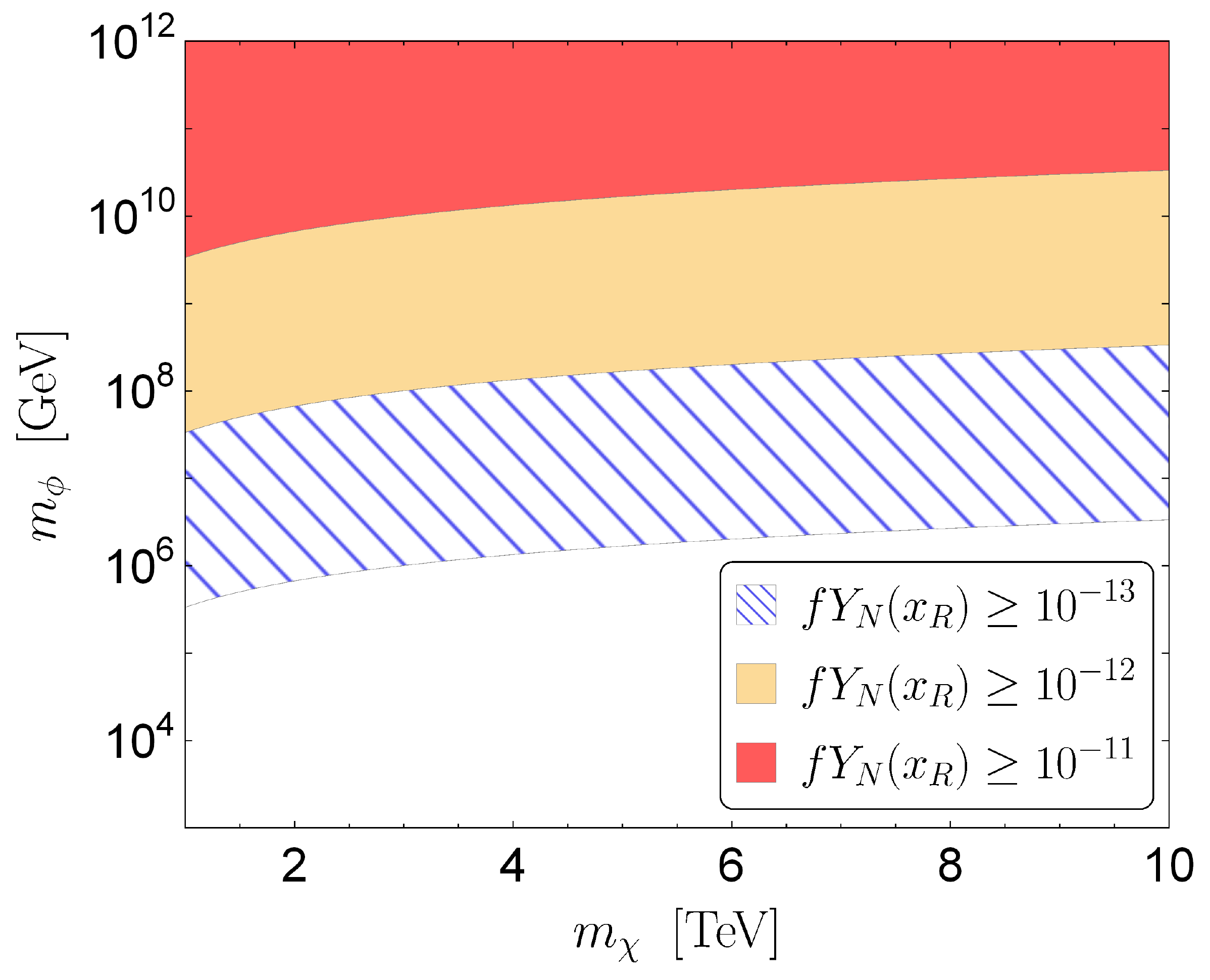}
\caption{% -----
The allowed region in $(m_{\chi} , m_{\phi})$ plane when the initial yield of the RH neutrino is given.
The initial yield is fixed as $f Y_{N}(x_R) \geq 10^{-13},\ 10^{-12},\ 10^{-11}$, and the small $\phi$ mass is favored.
}% -----
\label{fig:mphiTR}
\end{figure}
%%%%%
%

%
The time evolution of the yields are shown in Fig.~\ref{fig:nonthermal},
in which the masses are fixed as $m_{\phi} = 10^{6} \ \mathrm{GeV}$, $m_{\chi} = 10^{3}\ \mathrm{GeV}$ and $M_{N} = 10^{4} \ \mathrm{GeV}$.
The yield $Y_{N}$ initially created by the inflaton decay is large and remains the constant for $T \gtrsim m_{\phi}$, 
during which $\phi$ and Majoron are generated through the decay and scattering processes.
After the creation of Majoron DM by this process, the relic abundance is frozen-in at the temperature just below $m_{\phi}$.\footnote{
The relic abundance of the Majoron could be slightly changed by thermalized RH neutrinos. 
However, it is not large effect unless $m_\phi$ is close to $M_N$.}
On the other hand, 
the heavy scalar $\phi$ similarly created by the $N$ decay finally disappears after the $\phi \to \chi \chi$ process becomes effective in the thermal history.
In Fig.~\ref{fig:mphiTR}, we show the allowed region in $(m_{\chi}, m_{\phi})$ plane with the initial yield $ f Y_{N} (x_R) \geq 10^{-11},\ 10^{-12},\ 10^{-13}$.
A smaller $\phi$ mass is favored to realize the DM relic abundance.
The figure shows that a tiny value of the coupling $f$ is compatible with the observations, 
while that depends on the other parameters.
%

%%%%%%%%%%%%%%%
\section{Summary and outlook}
\label{052401_27Mar20}

We have studied the scenarios where the Majoron, a pNGB of lepton number symmetry with TeV-scale mass, 
can be the DM of the universe.
Since the decay constant of the Majoron is large and the coupling to the SM is tiny, 
it is nontrivial how to create the Majoron in the early universe, called Majorogenesis.
The Majoron model can realize neither
freeze-out nor freeze-in production of the Majoron DM with the large VEV
because the Majoron couplings to the SM particles are tiny 
and the Yukawa couplings to the RH neutrinos are flavor-diagonal
in the mass basis of the RH neutrinos.
To avoid this flaw, we have discussed three scenarios (A)--(C)
for Majorogenesis via the freeze-in mechanism;
(A) introducing explicit Majorana masses,
(B) using the interaction with the SM Higgs doublet,
(C) using the resonant production from the non-thermally induced RH neutrinos.
In (A), 
we find the lower bound on the Majoron Yukawa coupling for the freeze-in Majorogenesis to work, 
and the bound is roughly comparable with the tiny value of Yukawa coupling constrained from astrophysics.
Therefore, this scenario could be proved or excluded in the near future observations such as
Cherenkov Telescope Array (CTA) \cite{Carr:2015hta} and IceCube Neutrino Observatory \cite{Aartsen:2019swn}.
In (B), 
the toal coupling between the Majoron and the SM Higgs is found to be canceled and suppressed by the large mass scale, 
and is useful to create the Majoron via the freeze-in mechanism.
Note that this scenario is quite general because
we have used only the fact that $\chi$ is the pNGB having
the large VEV and the mixing coupling to the SM Higgs.
In (C), the sufficient amount of RH neutrinos are produced by the decay of the inflaton during the reheating.
After that, the $\phi$-mediated $NN\chi\chi$ interaction,
whose magnitude is constrained by cosmic-ray observations,
can be used to realize the freeze-in production.
In all the scenarios (A)--(C), 
there are the parameter regions realizing the DM relic abundance
and avoiding the astrophysical constraints.
Therefore, the Majoron with the TeV-scale mass (or heavier) can play the role of DM in the universe.
For further study, 
it may be intersting to examine the leptogenesis\cite{Fukugita:1986hr} in these scenarios.
A straightforward way is using resonances between the RH neutrinos \cite{Pilaftsis:2003gt}.
A more challenging is introducing other particles
whose masses are at an intermediate scale between $v$ and $v_\phi$.
One can use radiative decay processes of RH neutrinos where
the new particles appear in the loop to generate lepton asymmetry.
This motivates us to consider an extension 
in which one more SM-singlet $U(1)_L$-charged scalar is added.
Whether such type of leptogenesis can be compatible with the TeV-scale Majorogeneis is left for future work \cite{Abe:2020}.
%

%%%%%%%%%%%%%%%%%%%%%%%%%%%%%%%%%%%

\section*{Acknowledgments}
\noindent
The authors thank Motoko Fujiwara and Takashi Toma for useful discussions and comments.
The work is supported by JSPS Grant-in-Aid for Scientific Research,
No. JP20J11901 (Y.A.), No. JP18J22733 (Y.H.) and No. JP18H01214 (K.Y.).
%The work of Y.H. (K.Y.) is supported by JSPS Grant-in-Aid for Scientific Research, No. JP18J22733 (No. JP18H01214).
%The work of Y.A. is supported by JSPS Grant-in-Aid for Scientific Research, No. JP20J11901.
%
%%%%%%%%%%%%%%%%%%%%%%%%%%%%%%%%%%%

\end{document}